\pdfoutput=1
\documentclass[%
reprint,
superscriptaddress,
amsmath,amssymb,
aps,
prb,
]{revtex4-2}

\usepackage{graphicx}
\usepackage{dcolumn}
\usepackage{bm}
\usepackage{hyperref}

\usepackage{ulem}
\usepackage{xcolor}

\usepackage{comment}

\begin{document}
	\normalem
	\preprint{APS/123-QED}

	\title{Asymmetry-Based Quantum Backaction Suppression in Quadratic Optomechanics}
	
	\author{Vincent Dumont}
	\email{vincent.dumont@mail.mcgill.ca}
	\affiliation{Department of Physics, McGill University, Montr\'{e}al, Qu\'{e}bec H3A 2T8, Canada}
	
	\author{Hoi-Kwan Lau}
	\affiliation{Department of Physics, Simon Fraser University, Burnaby, British Columbia V5A 1S6, Canada}
	
	\author{Aashish A.~Clerk}
	\affiliation{Pritzker School of Molecular Engineering, University of Chicago, Chicago, Illinois 60637, USA}
	
	\author{Jack C.~Sankey}
	\affiliation{Department of Physics, McGill University, Montr\'{e}al, Qu\'{e}bec H3A 2T8, Canada}

	\date{\today}
	
	\begin{abstract}
		As the field of optomechanics advances, quadratic dispersive coupling (QDC) promise an increasingly feasible class of qualitatively new functionality. However, the leading QDC geometries also generate linear \textit{dissipative} coupling, and an associated quantum radiation force noise that is detrimental to QDC applications. Here, we propose a simple modification that dramatically reduces this noise \textit{without} altering the QDC strength. We identify optimal regimes of operation, and discuss advantages within the examples of optical levitation and nondestructive phonon measurement.
	\end{abstract}

	\maketitle

	\textit{Introduction.---} The field of optomechanics  \cite{Aspelmeyer2014Cavity} explores the forces exerted by light, and increasingly accesses the quantum regime with an eye toward sensing, information, and fundamental tests of macroscopic quantum motion. To date, the vast majority of experimental breakthroughs -- room-temperature generation of broadband squeezed light \cite{Aggarwal2020Room}, measurement near the Heisenberg limit or below the standard quantum limit \cite{Mason2019Continuous, Yu2020Quantum}, quantum information transduction 
	\cite{Chu2020A}, and creation of exotic quantum motion 
	\cite{Barzanjeh2022Optomechanics}, to list just a few -- have been achieved in systems having so-called linear dispersive coupling (LDC), wherein an optical resonance frequency depends linearly on the displacement of a mechanical element. As the field advances, systems exhibiting purely quadratic dispersive coupling (QDC), wherein the optical frequency depends on the \textit{square} of mechanical displacement, promise a wide and increasingly feasible range of applications that are qualitatively different, such as quantum nondemolition (QND) readout of phonon number \cite{Thompson2008Strong, Bhattacharya2008Optomechanical, Jayich2008Dispersive} or shot noise \cite{Clerk2010Quantum}, generation of exotic quantum states \cite{Vanner2011Selective, Nunnenkamp2010Cooling, Nunnenkamp2011Single} or entanglement \cite{Borkje2011Proposal}, non-reciprocal photon control \cite{Xu2020Quantum}, photon \cite{Rabl2011Photon} or phonon \cite{Xie2017Phonon} blockade, and stable center-of-mass \cite{Chang2010Cavity, Chang2012Ultrahigh} or torsional \cite{Muller2015Enhanced} optical traps for geometry tuned \cite{Barasheed2016Optically} and ultrahigh-Q mechanical systems.
	
	QDC optomechanical systems can mostly all be mapped onto the `membrane-in-the-middle' (MIM) \cite{Thompson2008Strong, Jayich2008Dispersive} paradigm, wherein a partially-reflective membrane splits a Fabry-Perot cavity into two identical sub-cavities, such that their LDC mutually cancels, producing purely QDC to leading order. This configuration has been successfully realized with a membrane in a macroscopic \cite{Thompson2008Strong} or microscopic \cite{Flowers2012Fiber} cavity, on-chip \cite{Paraiso2015Position, Grudinin2010Phonon}, levitated \cite{Bullier2020QuadraticARXIV}, or with atomic clouds \cite{Purdy2010Tunable}. Importantly, this quadratic \textit{dispersive} coupling is always accompanied by a linear \textit{dissipative} coupling 
	\cite{Elste2009Quantum, Sawadsky2015Observation, Weiss2013Quantum, Weiss2013Strong,Kilda2016Squeezed, Qu2015Generating,Gu2013Generation,Vyatchanin2016Quantum}, thereby introducing detrimental quantum radiation force noise (QRFN) \cite{Miao2009Standard,Yanay2016Quantum,Burgwal2020Comparing} that can place strict limits on what is possible, especially in the quantum regime \cite{Miao2009Standard, Yanay2016Quantum, Burgwal2020Comparing}. As such, understanding and mitigating this noise is of paramount importance.
	
	Here we propose an optomechanical geometry that can dramatically reduce QRFN \textit{without} compromising the strength of QDC. Specifically, our system exploits two non-identical sub-cavities, a situation that can be realized, e.g., by simply displacing the membrane in conventional MIM setups.  A previous classical wave analysis showed that this configuration can exhibit reduced linear dissipative coupling \cite{Dumont2019Flexure}, suggesting the possibility of suppressing QRFN in the quantum regime. Here we present a full quantum analysis to verify this conjecture and quantify its limits. We first derive the equations of motion, revealing the quantum mechanical origin of QDC, then quantify the QFRN with full consideration of fundamental sources of quantum noise.  In the ideal regime of negligible internal loss and a single-port cavity (i.e., where one mirror is perfectly reflective), we show that QFRN can be suppressed by many orders of magnitude by reducing the length of one sub-cavity. When internal loss is non-negligible, our analysis identifies the optimal configuration that minimizes QFRN, achieving more than two orders of magnitude suppression in a realistic system. We then discuss how this can be applied to cavity-assisted optical levitation, achieving noise below that of free-space traps (and MIM systems), and to improving the resolvability of QND phonon number readout.
	
	
	\textit{Quantum model.---}For illustrative purposes, we focus on the membrane-cavity geometry shown in Fig.~\ref{Fig1}(a); a detailed derivation including similar expressions for  general systems can be found in the supplementary materials \cite{supplementary}.
	Our setup consists of a cavity of length $L$ partitioned into two sub-cavities by a membrane with (field) transmission $|t_m|\ll 1$. Including the quantum mechanical displacement of the membrane $\hat{x}$, the length of the sub-cavities $1$ and $2$ are respectively $L_1 + \hat{x}$ and $L_2 - \hat{x}$, where $L_1$ and $L_2\equiv L-L_1$ are chosen such that the sub-cavity frequencies are degenerate at frequency $\omega_0 = N_1 \pi c/L_1 = N_2 \pi c/L_2$ for some integers $N_1$ and $N_2$. The Hamiltonian of the sub-cavity photonic fields is given by 
	$\hat{H}_\mathrm{opt} = \hat{H}_1 + \hat{H}_2 + \hat{H}_\mathrm{c}$, where, at the leading order of $\hat{x}$, the photonic energies are $\hat{H}_j = \hbar\left(\omega_{0}+ \hat{x}(-1)^j \omega_0/L_j\right) \hat{a}_{j}^{\dagger}\hat{a}_{j}$ for $j=1,2$, with $\hat{a}_j$ being the photon annihilation operator. The transmissive membrane couples the two sub-cavities via $\hat{H}_\text{c}=-\hbar J\left(\hat{a}_{1}^{\dagger}\hat{a}_{2}+\hat{a}_{2}^{\dagger}\hat{a}_{1}\right)$ at a rate $J = c|t_m|/(2\sqrt{L_1L_2})$ \cite{supplementary}.

	\begin{figure}
		\includegraphics[width=0.95\columnwidth]{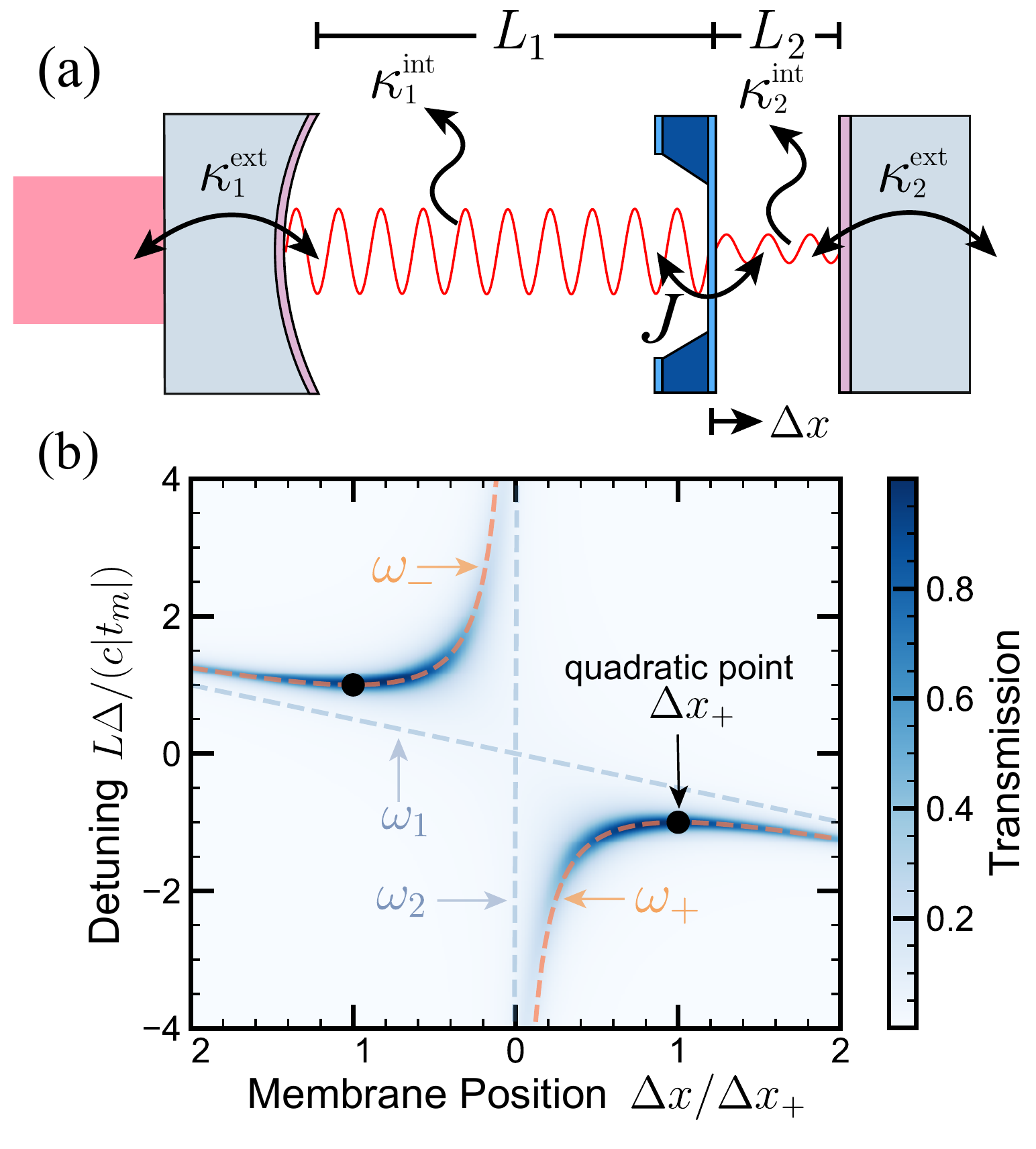}
		\caption{Membrane-cavity system. (a) Fabry-Perot cavity asymmetrically split by a membrane (blue) of transmission $|t_m|$ into two sub-cavities of lengths $L_j$. The sub-cavities are coupled to each other through the membrane at rate $J$.  Drive and detection are conducted through the external coupling at rates $\kappa^\textrm{ext}_j$, while photons are lost at rates $\kappa^\textrm{int}_j$. (b) Cavity transmission for varied membrane position $\Delta x$ 
			and drive detuning $\Delta \equiv \omega_\mathrm{in}-\omega_0$.
			In this example, $|t_m|^2 = 7\times 10^4$ ppm, mirror 1 transmission $|t_1|^2 = 6\times 10^4$ ppm, back mirror transmission $|t_2|^2 = 4\times 10^4$ ppm, and asymmetry $L_1/L_2 = 100$. Grey dashed lines show the uncoupled ($J=0$) sub-cavity frequencies $\omega_\textrm{in}=\omega_j \equiv \omega_0\left(1 + (-1)^j \Delta x/L_j\right)$; orange dashed lines show eigenfrequencies $\omega_\pm$.}
		\label{Fig1}
	\end{figure}

	In contrast to a MIM setup, the LDC of photonic eigenmodes is non-vanishing when $L_1 \neq L_2$. To obtain purely QDC, the mean position of the membrane needs to be slightly displaced to a \textit{quadratic point} (Fig.~\ref{Fig1}(b)). To illustrate the idea, we first define the membrane's quantum motion $\hat{z}$ around the classical displacement $\Delta x$, i.e. $\hat{x} \equiv \Delta x + \hat{z}$ where $\Delta x \ll L_j$. 
	The photonic Hamiltonian can be generally re-written as %
	\begin{equation}\label{eq:Hopt}
	\hat{H}_\textrm{opt} = \hat{H}_p + \hat{z}\hat{F}_\textrm{opt}~,
	\end{equation}
	where $\hat{H}_p$ involves only photonic fields, and 
	\begin{align}
	\hat{F}_\mathrm{opt} &= \hbar\omega_0 \left( \frac{1}{L_1} \hat{a}_1^\dagger \hat{a}_1 - \frac{1}{L_2} \hat{a}_2^\dagger \hat{a}_2 \right)
	\end{align}
	is the radiation force \footnote{The dependence of $J$ on membrane displacement is neglected here, because its effect is much smaller than that of the other optomechanical couplings.}. 
	
	At any $\Delta x$, the eigenmodes of $\hat{H}_p$ can be expressed succinctly as $\hat{a}_\pm (\Delta x) = \cos(\theta_\pm) \hat{a}_1  + \sin(\theta_\pm) \hat{a}_2$, where the amplitudes satisfy $\cot(2\theta_\pm) = \pm L \omega_0 \Delta x / \sqrt{L_1L_2}c|t_m|$. In terms of the eigenmodes, $\hat{H}_p = \hbar \omega_+ \hat{a}^\dag_+\hat{a}_+ + \hbar \omega_- \hat{a}^\dag_-\hat{a}_-$, with eigenfrequencies
	\begin{align}\label{eq:omegapm}
	\omega_{\pm}=\omega_{0} &+ \frac{(L_2-L_1)\omega_0 \Delta x}{2 L_1 L_2} 
	\mp\sqrt{\left(\frac{L\omega_0\Delta x}{2L_1L_2}\right)^{2}+ J^2} .
	\end{align}
	
	The radiation force can also be expressed in terms of the eigenmodes as $\hat{F}_\textrm{opt} = h_+ \hat{a}^\dag_+\hat{a}_+ + h_c (\hat{a}^\dag_+\hat{a}_- + \hat{a}^\dag_-\hat{a}_+) +  h_- \hat{a}^\dag_-\hat{a}_-$, where coefficients $h_+, h_c, h_-$ depend on the composition of eigenmodes, and thus $\Delta x$. The dispersive optomechanical coupling in Eq.~(\ref{eq:Hopt}) is generally linear (i.e., the adiabatic eigenfrequencies mostly depend linearly on position to leading order), with the exception of the ``quadratic points'' 
	%
	\begin{align}
	\Delta x_\pm =  \pm \frac{c |t_m|}{2\omega_0}  \frac{ L_1-L_2}{L }~,
	\end{align} 
	where the frequency of one eigenmode exhibits QDC to leading order in $\hat{z}$ (i.e., $h_\pm=0$ at $\Delta x_\pm$). For simplicity, we focus on $\Delta x_+$ hereafter, since the physics of interest is identical for $\Delta x_-$. The leading \textit{coherent} optomechanical effect on $\hat{a}_+$ will be a tunnelling with $\hat{a}_-$, but, due to the gap of eigenfrequencies in Eq.~(\ref{eq:omegapm}), $\hat{a}_-$ can be adiabatically eliminated, and the dynamics of $\hat{a}_+$ is governed by a pure QDC:
	\begin{equation}\label{eq:QDC}
	\hat{H}_+ 
	=\hbar \left(\omega_+ + \frac{2 \omega_0^2}{c |t_m| L} \hat{z}^2 \right) \hat{a}^\dag_+\hat{a}_+ ~.
	\end{equation}
	Importantly, our derivation shows that the overall QDC strength is determined by the cavity length $L$ but \textit{not} individual sub-cavity lengths $L_1$ and $L_2$, in agreement with classical calculations.

	
	\textit{Force noise from dissipative coupling.---}While the effective adiabatic Hamiltonian in Eq.~(\ref{eq:QDC}) is purely quadratic in $\hat{z}$, the full dynamics give rise to linear dissipative backaction. In particular, the cavity is coupled to an external environment for control and readout, and is also subjected to internal loss.  In the membrane-cavity setup, the external coupling and internal loss rates are, respectively,
	\begin{equation}\label{eq:kappa}
	\kappa^\textrm{ext}_j = c|t_j|^2/(2L_j)~~,~~\kappa^\textrm{int}_j=c\mathcal{T}_j/(2L_j)~,
	\end{equation}
	where $t_j$ is the end-mirror field transmission, and $\mathcal{T}_j$ is the round-trip internal photon loss fraction for each sub-cavity.  These environmental couplings create fluctuation in the sub-cavity fields that generates QRFN, with power spectral density $S_{FF}(\omega) = \int e^{i \omega t} \langle \delta \hat{F}(t) \delta \hat{F}(0)\rangle dt$ (where $ \delta \hat{F} \equiv \hat{F}_\textrm{opt}- \langle \hat{F}_\textrm{opt}\rangle$ is the optical force fluctuation \cite{Clerk2010Introduction}) we will now quantify.

	\textit{Ideal single-port cavity.---}To illustrate the potential for improvement, we first consider a single-port cavity ($\kappa^\text{int}_2 = \kappa^\text{ext}_2=0$).
	In the presence of a drive, $S_{FF}(\omega)$ can be calculated using the linearized Heisenberg-Langevin equations of motion \cite{Clerk2010Introduction}. If $\hat{a}_+$ mode is driven through mirror 1 at frequency $\omega_\mathrm{in}$, the QRFN power spectral density becomes
	\begin{align}\label{eq:force-noise-raw}
	S_{FF} (\omega) = & \hbar^2 |\bar{a}_+|^2 \kappa_1 L \left( \frac{\omega_0 }{L_1 L_2}\right)^2 \times \nonumber\\
	&   \frac{ \left| L_2  \tilde{\chi}_{11}(\omega) + i L_1 J\chi_2^*(0)\tilde{\chi}_{21}(\omega)  \right|^2}{\left| \sqrt{L_1} + i \sqrt{L_2} J \chi_2 (0) \right|^2}   ,
	\end{align}
	where $|\bar{a}_+|^2$ is the mean photon number in the $\hat{a}_+$ mode, $\kappa_j \equiv \kappa^\textrm{ext}_j+ \kappa^\textrm{int}_j$ is the total loss rate of sub-cavity $j$, $\chi_j(\omega) = [-i(\omega_\mathrm{in}-\omega_0 +  (-1)^{j+1} \omega_0 \Delta x_+/L_j + \omega) + \kappa_j/2]^{-1}$ is the susceptibility,
	and $\tilde{\chi}_{lm}$ are matrix elements of the eigenmode susceptibility
	\begin{align}
	\boldsymbol{\tilde{\chi}} (\omega) &=\frac{1}{\chi_1^{-1}(\omega)\chi_2^{-1}(\omega)+J^2}\begin{pmatrix}
	\chi_2^{-1}(\omega) & iJ \\
	iJ & \chi_1^{-1}(\omega) 
	\end{pmatrix}.
	\end{align}
	In the `large-gap' limit $2c|t_m|/L \gg\kappa_1$, $\kappa_2$, $|\omega|$, and when the drive is resonant with $\hat{a}_+$ ($\omega_\mathrm{in} = \omega_+$),
	Eq.~(\ref{eq:force-noise-raw}) reduces to
	\begin{align}
	S_{FF} (\omega) &= \label{eq:force-noise-single-port-membrane} \left(\frac{2L_2}{L}\right)^{2}\frac{\hbar^{2}|\bar{a}_{+}|^{2}\omega_{0}^{2}}{c^{2}|t_{m}|^{2}}\frac{\omega^2\kappa_{+} }{\omega^2+\kappa_{+}^{2}/4}
	\end{align}
	where the
	$\hat{a}_+$ decay rate $\kappa_+ (\Delta x_+) = \kappa_1 L_1/L$.
	The advantage of our setup is now clear:
	when the membrane is positioned near the back mirror, ($L_2 \ll L_1 \approx L$),
	QRFN is suppressed by a factor $(2L_2/L)^2$. In a centimeter-scale cavity with wavelength-scale membrane-mirror separation \cite{Dumont2019Flexure}, this suppression factor can reach $(2L_2/L)^2 \sim 10^{-8}$ when compared to the MIM geometry. We stress that the quadratic coupling in Eq.~(\ref{eq:QDC}) remains unchanged in this limit.
	

	\begin{figure} 
		\includegraphics[width=0.95\columnwidth]{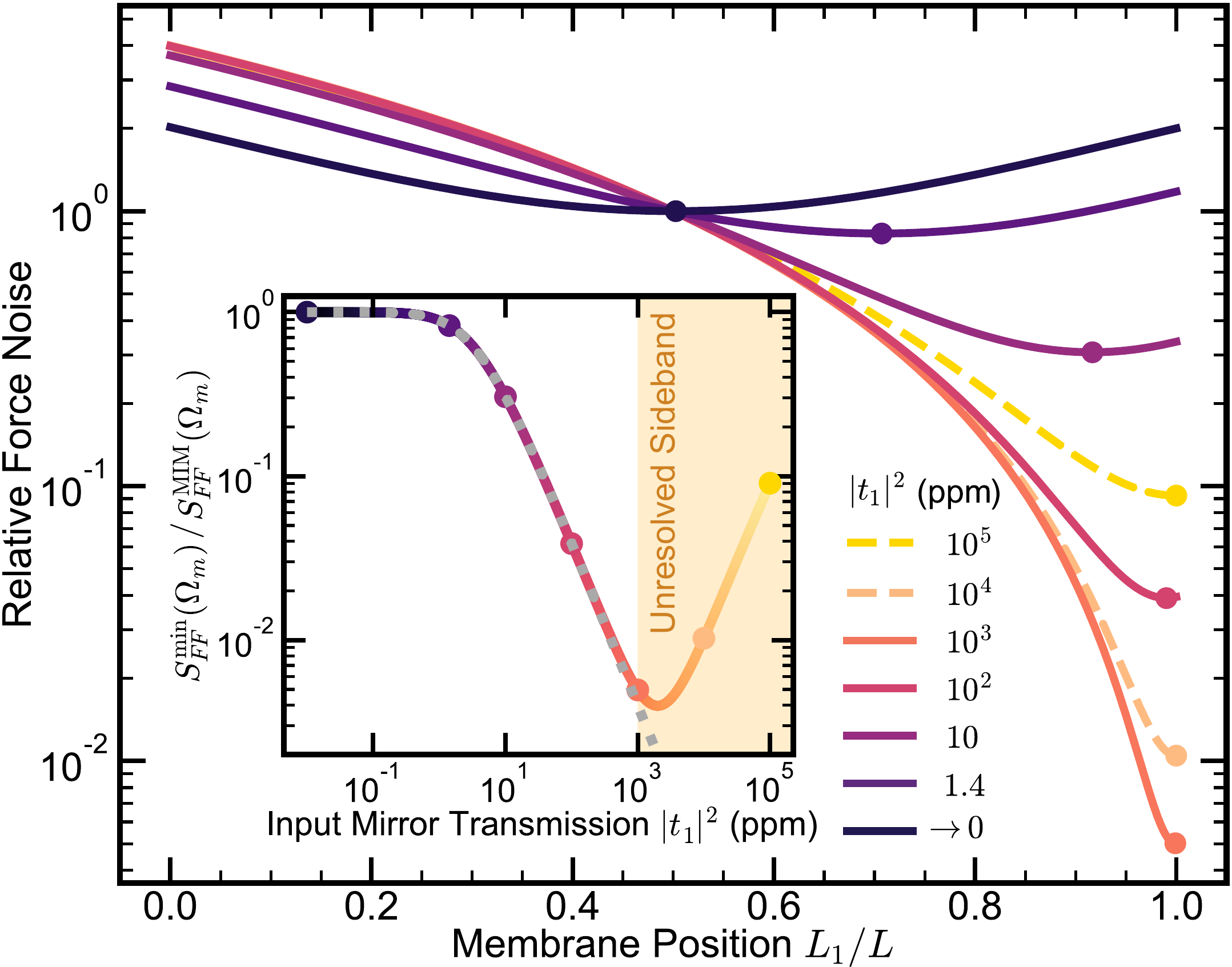}
		\caption{Force noise $S_{FF}(\Omega_m)$ for varied configurations.
			For each set of parameters, the force noise is normalized by the MIM value, i.e. when $L_1=L/2$.
			System parameters correspond to a ``typical'' optomechanical setup: 
			cavity length $L=10$~cm, membrane transmission $|t_m|^2 = 10^{4}$ ppm \cite{Stambaugh2014From}, $|t_2|^2=0$, mirror loss $\mathcal{T}_j = 1$ ppm \cite{Rempe1992Measurement, Kong2021PhysicalARXIV}, and mechanical frequency $\Omega_m = 2\pi \times 240$ kHz, with a wide range of input transmissions $|t_1|^2$. The solid (dashed) curves are in the resolved (unresolved) sideband regime, i.e. $\Omega_m > \kappa_+$ ($\Omega_m < \kappa_+$). The markers indicates the optimal membrane position derived in Eq.~(\ref{eq:x_min}). Inset: optimal reduction of force noise relative to the MIM setup. The orange shaded area indicates the unresolved sideband regime, and the dotted line corresponds to the resolved-sideband approximation in Eq.~(\ref{eq:punchline}).}
		\label{Fig2} 
	\end{figure}
	
	\textit{Lossy cavity.---}Practical photonic resonators suffer from internal losses that play an important role. While loss in sub-cavity 1 simply rescales $\kappa_1$ in Eq.~(\ref{eq:force-noise-single-port-membrane}), the loss in sub-cavity 2 (i.e. $\kappa_2 \neq 0$) leads to QRFN spectral density
	\begin{align}
	S_{FF}(\omega) 	= \frac{\hbar^{2}|\bar{a}_{+}|^{2}\omega_{0}^{2}}{c^{2}|t_{m}|^{2}}  \frac{ L_{2}}{L^{2}}  \frac{4L_{1}\kappa_{-}\omega^{2}+L\kappa_{+}^{2}\kappa_{2}}{\omega^{2}+\kappa_{+}^{2}/4}\label{eq:force-noise-large-gap-membrane} ~,
	\end{align}
	where the dissipation rate of $\hat{a}_+$ and $\hat{a}_-$ modes are, respectively, $\kappa_+ = \kappa_1 L_1/L  + \kappa_2 L_2/L$ and $\kappa_- = \kappa_1 L_2/L  + \kappa_2 L_1/L$.

	In a modified form of the resolved-sideband regime, wherein the mechanical frequency $\Omega_m \gg \kappa_+ \sqrt{L\kappa_2/4L_1\kappa_-}$ \cite{supplementary}, the force noise in Eq.~(\ref{eq:force-noise-large-gap-membrane}) then simplifies to
	\begin{align}
	S_{FF}(\Omega_m) 	&= 4 \frac{\hbar^{2}|\bar{a}_{+}|^{2}\omega_{0}^{2}}{c^{2}|t_{m}|^{2}}  \frac{  L_1 L_{2}}{L^{2}} \kappa_{-}  .
	\end{align}
	Because $\kappa_2$ scales inversely with $L_2$ (due to the changing round-trip time) in Eq.~(\ref{eq:kappa}), force noise cannot be suppressed indefinitely by shrinking the second sub-cavity. Instead, $S_{FF}(\Omega_M)$ reaches a minimum at an optimal first sub-cavity length
	\begin{align}\label{eq:x_min}
	L_{1, \mathrm{min}} = \frac{|t_1|^2 + \mathcal{T}_1 }{|t_1|^2 + \mathcal{T}_1 + |t_2|^2 + \mathcal{T}_2}L~.
	\end{align} 
	Notably, this is also where the sub-cavities have equal dissipation ($\kappa_1 = \kappa_2$). At quadratic points near this optimal position, 
	\begin{align} \label{eq:Sff_min}
	S_{FF}^{\mathrm{min}}(\Omega_m) 	&   	=2\frac{\hbar^{2}|\bar{a}_{+}|^{2}\omega_{0}^{2}}{cL|t_{m}|^{2}} \frac{(|t_1|^2 + \mathcal{T}_1)( |t_2|^2 + \mathcal{T}_2)}{|t_1|^2 + \mathcal{T}_1 + |t_2|^2 + \mathcal{T}_2} .
	\end{align}
	When comparing with the force noise of the MIM setup ($L_1 = L/2$), the noise is then suppressed by a factor
	\begin{align}\label{eq:punchline}
	\frac{S_{FF}^\mathrm{min} }{ S_{FF}^{\mathrm{MIM}}	}  &=4  \frac{ (|t_1|^2 + \mathcal{T}_1)( |t_2|^2 + \mathcal{T}_2)}{(|t_1|^2 + \mathcal{T}_1 + |t_2|^2 + \mathcal{T}_2)^2 }\rightarrow   \frac{4  \mathcal{T}_2}{|t_1|^2},
	\end{align}
	where the last step is in the typical near-single-port limit $|t_1|^2 \gg \mathcal{T}_1, \mathcal{T}_2, |t_2|^2$ \footnote{Note we consider only experiments where the signal is collected from one port, so the non-vanishing transmission of the end mirror can be regarded as additional internal loss.}. Eq.~(\ref{eq:punchline}) is our main result: even with internal loss, QRFN can be significantly suppressed by simply placing the mechanical mirror away from the mid-point, \textit{without} reducing the QDC strength (c.f. Eq.~(\ref{eq:QDC})).
	
	For completeness, we note that the force noise in Eq.~(\ref{eq:force-noise-large-gap-membrane}) can also be minimized outside the resolved sideband regime, yielding
	\begin{equation}
	S_{FF}^\mathrm{min} (\Omega_m)
	=2 \frac{\hbar^{2}|\bar{a}_{+}|^{2}\omega_{0}^{2}}{cL|t_{m}|^{2}}    (|t_2|^2 + \mathcal{T}_2) \frac{1 +4\mathcal{B}\Omega_m^2/\kappa_+^2}{1+4 \Omega_m^2/\kappa_+^2},
	\end{equation}
	where $\mathcal{B} \equiv (|t_1|^2 + \mathcal{T}_1)/(|t_1|^2 + \mathcal{T}_1 + |t_2|^2 + \mathcal{T}_2)$.
	%
	
	To get a sense of the practical performance of our scheme in realistic laser cavities with dielectric mirrors,
	Fig.~\ref{Fig2} shows these results for a variety of parameters. Importantly, this setup can suppress QRFN by more than two orders of magnitude relative to the conventional MIM configuration. Our scheme is most advantageous in the single-port limit, with diminishing returns as the system enters the deeply unresolved sideband regime. 
	
	This main result directly benefits all QDC applications requiring low force noise; in the remaining text, we discuss this within the illustrative examples of optical levitation and QND phonon measurement.
	%

	\textit{Optical levitation.---}
	By placing a minimally supported reflector inside a cavity at a quadratic point, the reflector's motion collinear with the cavity axis can be optically trapped. For a resolved-sideband cavity driven at frequency $ \omega_\mathrm{in}=\omega_+\approx\omega_0$, the dispersive optical spring constant at $\Delta x_+$ is
	\begin{align}
	\hbar \omega'' |\bar{a}_+|^2 \approx  \frac{8}{|t_m|} \frac{\omega_\mathrm{in} \bar{P}_\mathrm{circ}}{c^2}~,
	\end{align}
	where $\omega''/2 = 2 \omega_0^2/c |t_m| L$ is the QDC strength in Eq.~(\ref{eq:QDC}). This spring constant is identical to that of a standing wave in free space for the same circulating power $\bar{P}_\mathrm{circ}$. However, a free space trap's QRFN $S_{FF}^\mathrm{FS} \approx 8\hbar \omega_\mathrm{in} \bar{P}_\mathrm{in}/c^2$, meaning our optimal membrane-cavity system has a relative force noise (from Eq.~\ref{eq:Sff_min})
	\begin{align}
	\frac{S_{FF}^{\mathrm{min}}(\Omega_m)}{S_{FF}^\mathrm{FS}} 	&=   \frac{1}{2|t_m|^2} \frac{(|t_1|^2 + \mathcal{T}_1)( |t_2|^2 + \mathcal{T}_2)}{|t_1|^2 + \mathcal{T}_1 + |t_2|^2 + \mathcal{T}_2} \\
	&\rightarrow  \frac{|t_2|^2 + \mathcal{T}_2}{2|t_m|^2},  \label{eq:minimum-trap-force-noise}
	\end{align}
	where the last expression is in the nearly-single-port limit. This means that, as long as most of the back-cavity light leaves through the membrane (i.e. $|t_m|^2 \gg |t_2|^2, \mathcal{T}_2$), QRFN can be significantly suppressed relative to free space (and MIM, as per Fig.~\ref{Fig2} and Eq.~(\ref{eq:punchline})). Furthermore, $\bar{P}_\text{circ}$ in a cavity system is achieved with much less input power, making it far easier to realize a quantum-limited light source that actually reaches this limit.
	
	\textit{QND phonon measurement.---}
	In the resolved-sideband regime, QDC naturally measures the time-averaged mechanical energy, enabling quantum nondemolition (QND) readout of phonon number \cite{Thompson2008Strong, Jayich2008Dispersive, supplementary}, which is usually proposed assuming a near-single-port cavity. We quantify the quantum-limited performance of such measurements with the ratio of measurement rate $\Gamma_\mathrm{meas}$ to backaction rate $\Gamma_{\mathrm{ba},n}$ \cite{Miao2009Standard}, a figure of merit that exceeds one when it is possible to resolve phonon number state $n$ before QRFN causes a jump.
	For our setup, when $\hat{a}_+$ mode is driven on resonance, this ratio becomes  (again assuming the large-gap limit, and a modified resolved-sideband limit $\Omega_m/\kappa_+ \gg \sqrt{(g_1+g_2)\kappa_2/(4g_2\kappa_-)} $ \cite{supplementary})
	\begin{align}
	\frac{\Gamma_\mathrm{meas}}{\Gamma_{\mathrm{ba},n}} 
	&= 
	\frac{64}{2n+1}\left[\frac{g_1 g_2}{\kappa_- \kappa_+}\right] \left(\frac{\kappa^\textrm{ext}_1}{\kappa_+}\frac{L_1}{L}\right) \equiv   \frac{x_\mathrm{zpf}^2 }{  x_\mathrm{res}^2  }  \label{eq:qnd-ratio-general},
	\end{align}
	where $g_j = \omega_0 x_\textrm{zpf}/L_j$ is the single-photon optomechanical coupling rate for sub-cavity $j$, and $x_\mathrm{zpf} = \sqrt{\hbar/2m\Omega_m}$ is the zero-point fluctuation of the membrane (mass $m$). We also define a ``resolvable variance'' $x_\mathrm{res}^2$, which is the value of $x_\textrm{zpf}^2$ the membrane must have in order to achieve unity $\Gamma_\text{meas}/\Gamma_{\text{ba},n}$. 
	
	The parenthetical factor $\kappa^\textrm{ext}_1L_1/\kappa_+L$ captures the cavity mode's input coupling efficiency, while the bracketed factor $\sqrt{g_1 g_2/\kappa_-\kappa_+}$ characterizes the single-photon strong coupling in our asymmetric system. 
	
\begin{figure}  
		\includegraphics[width=0.9\columnwidth]{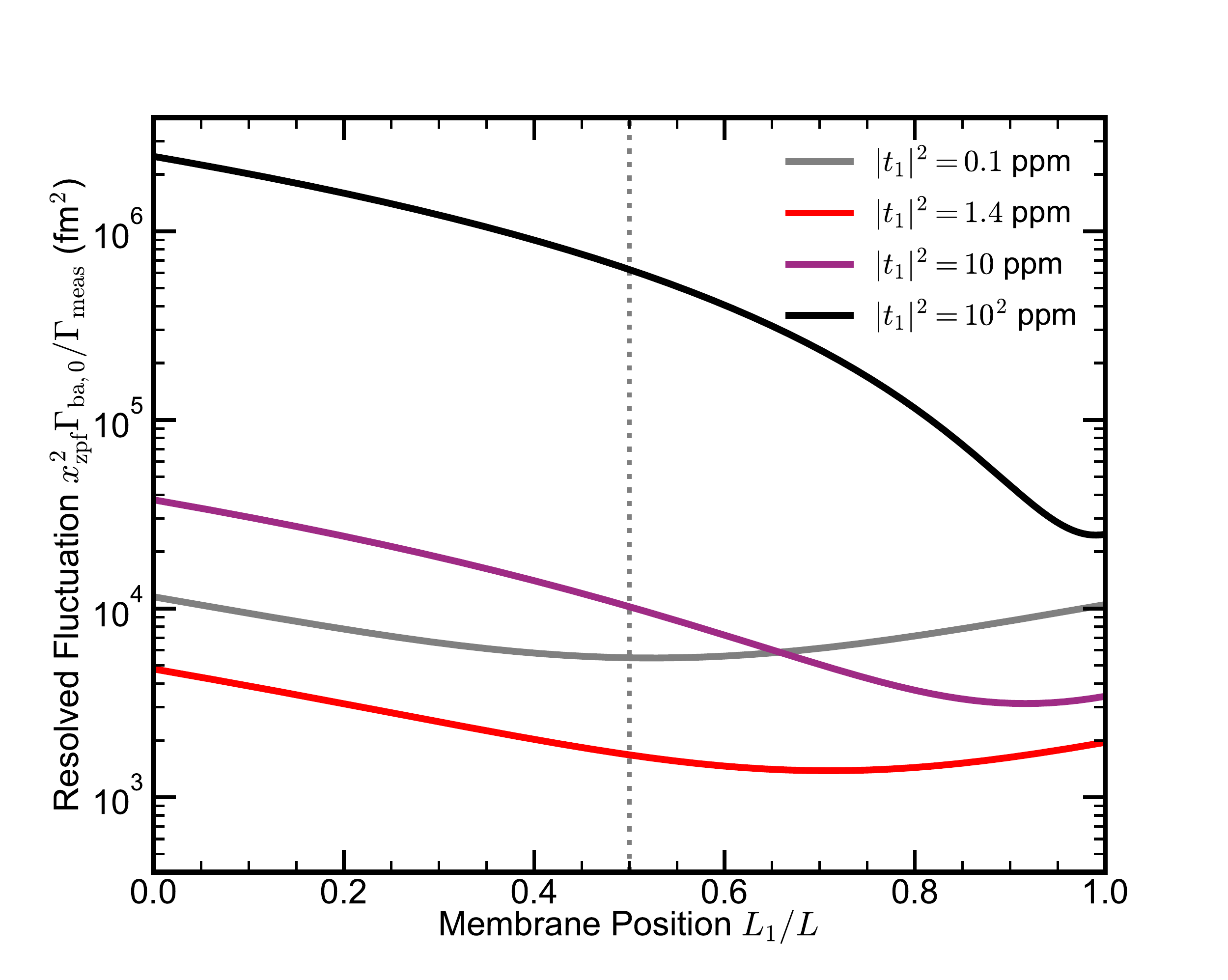}
		\caption{Resolvable fluctuation $x_\mathrm{res}^2$ for QND phonon measurement of the ground state ($n=0$).  	Our scheme exhibits clear advantage for a wide range of front mirror transmission $|t_1|^2$ (black and purple), and provides additional improvement when optimizing $|t_1|^2_\textrm{opt}$ (red). Further reducing $|t_1|^2$ provides no benefit due to reduced collection efficiency. In all cases, the system parameters are $\mathcal{T}_1 = \mathcal{T}_2 = 1$ ppm, $|t_2|^2 = 0$ ppm (a loss-limited Bragg stack), $|t_m|^2=10^4$ ppm, and $L = 10$ cm, $\omega_\textrm{in}=\omega_+$ in the resolved-sideband limit. The dotted vertical line indicates $L_1 = L/2$. 
		} 
		\label{Fig3} 
	\end{figure}

	For single-port approaches, Eq.~(\ref{eq:qnd-ratio-general}) yields dramatic improvements over the MIM approach, as shown in Fig.~\ref{Fig3}. In fact, this improvement is again the same large ratio that controls the force noise suppression in Eq.~(\ref{eq:punchline}); this is not surprising, since $\Gamma_{\mathrm{ba}, n} \propto S_{FF}$, and the quadratic coupling strength is independent of position. 
	We emphasize the practical benefits of performing these measurements with larger $|t_1|^2$, notably boosting the collected signal above detector noise and easing the laser-lock process. In systems operating far from the single-port regime, our analysis also identifies an optimal input mirror transmission \cite{supplementary}
	\begin{align}\label{eq:optimized_T1}
	|t_1|^2_\textrm{opt} = \sqrt{\mathcal{T}_1 (\mathcal{T}_1 + \mathcal{T}_2 + |t_2|^2)}.
	\end{align}
	that is neither single-port nor balanced. This further improves $x_\text{res}^2$ as shown by the red curve in Fig.~\ref{Fig3}.

	\textit{Summary.---}We propose an optomechanical setup that dramatically reduces quantum radiation force noise without affecting the quadratic dispersive coupling strength. For the illustrative membrane-cavity geometry, this is implemented by simply relocating the membrane toward the mirror with higher reflectivity. Our full quantum analysis identifies optimal configurations, and we demonstrate its advantage in optical levitation and nondemolition phonon measurement. Owing to its ease of implementation and our universal desire to control noise, we expect this proposal will immediately impact all optomechanical experiments aiming to exploit quadratic dispersive coupling, moving forward.
	
	We thank Simon Bernard, Lilian Childress, Bill Coish,  Zo\'{e} McIntyre, Christoph Reinhardt and  Yariv Yanay for fruitful discussions. VD acknowledges financial support from FRQNT-B2 Scholarship and McGill Schulich Graduate Fellowship.  HKL acknowledges support from Canada Research Chairs. AC acknowledges support from the Simons Foundation under a Simons Investigator Award. JCS acknowledges relevant support from the Natural Sciences and Engineering Research Council of Canada (NSERC RGPIN 2018-05635), Canada Research Chairs (235060), Institut Transdisciplinaire d'Information Quantique (INTRIQ), and the Centre for the Physics of Materials (CPM) at McGill.

	\addcontentsline{toc}{section}{References}
	\bibliographystyle{bibstyle-jack}
	\bibliography{errthing}
	
	\pagebreak
	\appendix
	
	\onecolumngrid

	\huge \begin{center} Supplementary Information \end{center}\normalsize
	
	\section{Backaction Model} \label{app:General_Model}

	Here we derive the expressions for the quantum radiation pressure force noise (QRFN) in an asymmetric cavity optomechanical system. We begin with the general optical equations of motion, eigenmodes, and quadratic coupling in Sec.~\ref{app:A1}, then focus on geometries having purely quadratic dispersive coupling in Sec.~\ref{app:A2}, showing that asymmetry can lead to a large reduction of force noise in Sec.~\ref{app:Force_Noise_Reduction} at the optimal membrane position. Finally, we discuss quadratic optical trapping (Sec.~\ref{app:Optical_Trapping}), and quantum non-demolition (QND) mechanical energy measurement Sec.~\ref{app:A3}, both of which benefit from reduced force noise.

	
	\subsection{Equations of Motion, Eigenmodes, Dissipation and Optomechanical Coupling}\label{app:A1}
	
	\textbf{General Equations of Motion:}
	To model the optical dynamics, we first write down the input-output equations of motion \cite{Clerk2010Introduction} for sub-cavity operators $\hat{a}_1$ and $\hat{a}_2$ in the frame rotating at an external drive frequency $\omega_\mathrm{in}$. These are
	\begin{align}\label{eq:EOM}
	\begin{pmatrix}
	\dot{\hat{a}}_1 \\
	\dot{\hat{a}}_2 
	\end{pmatrix}
	&= -i\boldsymbol{\tilde{H}}  	\begin{pmatrix}
	\hat{a}_1 \\
	\hat{a}_2
	\end{pmatrix} 
	+
	\boldsymbol{\tilde{\kappa}^\text{ext}}
	\begin{pmatrix}
	\hat{a}_{1}^\mathrm{in}\\
	\hat{a}_{2}^\mathrm{in}
	\end{pmatrix}
	+
	\boldsymbol{\tilde{\kappa}^\text{int}}
	\begin{pmatrix}
	\delta\hat{a}_{1}^\mathrm{int}\\
	\delta\hat{a}_{2}^\mathrm{int}
	\end{pmatrix}\\
	\boldsymbol{\tilde{H}}&\equiv 
	\begin{pmatrix}
	-(\Delta - G_1 \Delta x) -i \kappa_1/2 & -J\\
	-J & -(\Delta - G_2 \Delta x) -i\kappa_2/2 
	\end{pmatrix}\\
	\boldsymbol{\tilde{\kappa}_\text{ext}}
	&\equiv\begin{pmatrix}
	\sqrt{\kappa_{1}^\mathrm{ext}} & 0 \\
	0 & \sqrt{\kappa_{2}^\mathrm{ext}} 
	\end{pmatrix}\\
	\boldsymbol{\tilde{\kappa}_\text{int}}
	&\equiv \begin{pmatrix}
	\sqrt{\kappa_{1}^\mathrm{int}} & 0 \\
	0 & \sqrt{\kappa^{\mathrm{int}}_2}
	\end{pmatrix},
	\end{align}
	where $\hat{a}_{j}^\text{in}$ is the external drive applied to sub-cavity $j$, $\kappa_{j}^\text{ext}$ is the associated (power) coupling rate,  $\kappa_{j}^\mathrm{int}$ is the sub-cavity's internal loss rate, $\kappa_j = \kappa_{j}^\text{ext} + \kappa_{j}^\mathrm{int} $ is the total sub-cavity decay rate, $G_j$ is the sub-cavity's optomechanical coupling,  and $\Delta = \omega_\text{in}-\omega_0$ is the drive's detuning relative to the frequency $\omega_0$ at which modes $\hat{a}_1$ and $\hat{a}_2$ are degenerate, which we define to occur at mechanical displacement $\Delta x\equiv 0$.
	
	
	\textbf{Optical Eigenmodes:}
	The system's eigenfrequencies
	\begin{align}\label{eq:eigenfrequencies}
	\omega_{\pm}(\Delta x) =\omega_{0} &+ \frac{G_2 + G_1}{2} \Delta x  \mp\sqrt{\left(\frac{G_2 - G_1}{2}\Delta x\right)^{2}+J^{2}},
	\end{align}
	are obtained by diagonalizing the Heisenberg equations of motion ($\dot{\hat{a}}_j = -\frac{i}{\hbar} [\hat{a}, \hat{H}_\mathrm{opt}]$), and have associated eigenmodes that can be written succinctly as
	\begin{align}\label{eq:eigenmodes}
	\hat{a}_\pm (\Delta x) &= \hat{a}_1 \cos(\theta_\pm[\Delta x]) + \hat{a}_2 \sin(\theta_\pm[\Delta x]),
	\end{align}
	where the amplitudes satisfy
	\begin{align}
	\cot(2\theta_\pm[\Delta x])  &  = \pm\frac{G_2 - G_1}{2 J}\Delta x.
	\end{align} 

	\textbf{Dispersive Couplings:} When $G_1\ne G_2$, the eigenfrequencies of Eq.~\ref{eq:eigenfrequencies} exhibit an avoided crossing structure, with linear dispersive coupling (LDC)
	\begin{align}
	\frac{\partial \omega_\pm ( \Delta x )}{\partial x} =& \frac{G_1 + G_2}{2} \mp\frac{(G_1-G_2)^2}{2\sqrt{4J^2 + \left[(G_1 - G_2)\Delta x \right]^2}} \Delta x   
	\end{align}
	which, if $G_1<0$ and $G_2>0$, becomes zero at ``quadratic points''
	\begin{align} 
	\Delta x_\pm &=  \pm \frac{J (G_2+ G_1)}{(G_2 - G_1) \sqrt{-G_1G_2}}.
	\end{align}
	The frequencies at these extrema are
	\begin{align}
	\omega_\pm(\Delta x_\pm)& = \omega_0 \mp 2J \frac{\sqrt{-G_1 G_2}}{G_2 - G_1},
	\end{align}
	corresponding to an avoided gap $4J\sqrt{-G_1G_2}/(G_2-G_1)$. The quadratic dispersive coupling is generally
	\begin{align}
	\frac{\partial^2 \omega_\pm(\Delta x)}{\partial x^2} = \mp \frac{2J^2(G_1 - G_2)^2}{\left(4J^2 + \left[(G_1 - G_2)\Delta x \right]^2 \right)^{3/2}},
	\end{align} 
	which, at $\Delta x_\pm$, simplifies to
	\begin{align}
	\left. \frac{\partial^2 \omega_\pm}{\partial x^2}\right |_{\Delta x=\Delta x_\pm} &= \mp 2 \frac{1}{J} \frac{(-G_1 G_2)^{3/2}}{G_2 - G_1}\label{eq:d2omega_+-dx^2}.
	\end{align}
	The expressions (as with others involving $G_j$ and $J$ in this document) still represent a general system with two optical modes linearly coupled to one mechanical mode. For a membrane in a cavity, where $G_j = (-1)^j \omega_0/L_j$ and $J=c|t_m|/2\sqrt{L_1L_2}$ (see Appendix \ref{app:HoppingRate}), with sub-cavity length $L_j$ and membrane (amplitude) transmission $t_m$, this becomes
	\begin{align}
	\left. \frac{\partial^2 \omega_\pm}{\partial x^2}\right |_{\Delta x=\Delta x_\pm} &= \mp \frac{4}{|t_m|} \frac{\omega_0^2}{cL} \label{eq:d2omega_+-dx^2_membrane},
	\end{align}
	where $L= L_1 + L_2$ is the total cavity length.
	
	\textbf{Eigenmode Decay Rates:}
	Combined with the sub-cavity (power) losses $\kappa_j = \kappa_{j}^\mathrm{int} + \kappa_{j}^\mathrm{ext}$ (note these exclude $J$), the amplitudes attached to $\hat{a}_1$ and $\hat{a}_2$ in Eq.~\ref{eq:eigenmodes} permit the calculation of power decay rates
	\begin{align}
	\kappa_\pm (\Delta x) & = \kappa_1 \cos^2(\theta_\pm[\Delta x])  +  \kappa_2 \sin^2(\theta_\pm[\Delta x]).\label{eq:kappa_pm}  \\
	& = \bar{\kappa} \mp  \frac{( G_1 - G_2) \Delta x}{\sqrt{4J^2 +  \left[(G_2-G_1) \Delta x\right]^2}} \Delta \kappa,
	\end{align}
	for eigenmodes $\hat{a}_\pm$, where $\bar{\kappa} \equiv (\kappa_1+ \kappa_2)/2$ and $\Delta \kappa  \equiv (\kappa_1 - \kappa_2)/2$. At the quadratic points,
	\begin{align}
	\kappa_\pm(\Delta x_+) &= \bar{\kappa} \pm \Delta \kappa \frac{G_2 + G_1}{G_2 - G_1}\\
	\kappa_\pm(\Delta x_-) &= \bar{\kappa} \mp \Delta \kappa \frac{G_2 + G_1}{G_2 - G_1}.
	\end{align}
	For the case of a membrane-cavity system, we can write these (choosing $\Delta x_+$) in terms of the end mirror power transmissions $|t_j|^2$ and sub-cavity round-trip power losses $\mathcal{T}_j$:
	\begin{align}
	\kappa_+ (\Delta x_+)   & = \frac{c}{2L}  (|t_1|^2 + \mathcal{T}_1 + |t_2|^2 + \mathcal{T}_2) \equiv \kappa_0 \label{eq:kappa0} \\
	\kappa_- (\Delta x_+)   & =  \frac{c}{2L} \frac{L_2^2  (|t_1|^2 + \mathcal{T}_1)   + L_1^2 (|t_2|^2 + \mathcal{T}_2)} {L_1L_2}\label{eq:kappa_-_MIC},
	\end{align}
	where we have used $\kappa_{j}^\mathrm{int} = c\mathcal{T}_j/2L_j$ and $\kappa_{j}^\mathrm{ext} = c|t_j|^2/2L_j$. At this location, $\kappa_+$ is identical to the empty cavity decay rate $\kappa_0$ due to the fact that sub-cavity circulating powers are identical. This is true for any quadratic point we choose. 
	
	\textbf{Dissipative Coupling:} From Eq.~\ref{eq:kappa_pm}, the linear \textit{dissipative} coupling is  
	\begin{align}
	\partial_x \kappa_\pm(\Delta x) = \mp 4 J^2 \frac{( G_1 - G_2)\Delta \kappa }{\left( 4J^2 +  \left[(G_2-G_1) \Delta x\right]^2\right)^{3/2}} ,
	\end{align} 
	which, at the quadratic point, becomes
	\begin{align}
	\partial_x \kappa_\pm(\Delta x_\pm)  &= \pm 4 \frac{\Delta \kappa}{J} \frac{(-G_1 G_2)^{3/2}}{(G_2 - G_1)^2},
	\end{align}
	with corresponding single-photon strong coupling parameter
	\begin{align} \label{eq:DissipativeCouplingB}
	\tilde{B}_\pm (\Delta x_\pm)&  \equiv x_\mathrm{zpf} \left. \frac{\partial_x \kappa_\pm }{\kappa_\pm}\right|_{\Delta x = \Delta x_\pm} =    \pm 2 \frac{ 1}{J} \frac{(-G_1 G_2)^{3/2}}{G_2(G_2 - G_1)}   x_\mathrm{zpf} 
	\end{align} 
	in the single-port limit ($\kappa_2\rightarrow 0$). We can now see that this coupling can be greatly reduced in the limit $|G_2| \gg |G_1|$. For a single-port membrane-cavity, e.g., %
	\begin{align}
	\tilde{B}_\pm (\Delta x_\pm)&  \rightarrow    \pm  \frac{ 4}{|t_m|} \frac{\omega_0}{c}  \frac{L_2}{ L}   x_\mathrm{zpf},
	\end{align}
	which diminishes as $L_2/L$, suggesting a commensurate reduction of QRFN achieved by simply moving the membrane toward the back mirror.
	

	\textbf{Equations of Motion at Quadratic Point:}
	Finally, we write down the equations of motion for the eigenmodes $\hat{a}_\pm$ themselves at a quadratic point $\Delta x = \Delta x_+$. First, defining 
	\begin{align}
	\alpha & \equiv \sin(\theta_+ [\Delta x_+]) = \cos(\theta_-[\Delta x_+]) =  (1-G_2/G_1)^{-1/2}\label{eq:alpha}\\
	\beta  & \equiv \cos(\theta_+ [\Delta x_+]) = -\sin(\theta_-[\Delta x_+]) = (1-G_1/G_2)^{-1/2}\label{eq:beta}. 
	\end{align}
	for convenience, the equations of motion in the eigenmode basis (Eqs.~\ref{eq:eigenmodes}) become
	\begin{align}
	\dot{\hat{a}}_+	=&  -\left(-i \delta_+   + \frac{\kappa_1}{2} \beta^2 + \frac{\kappa_2}{2} \alpha^2  \right)  \hat{a}_+  - \alpha\beta \Delta \kappa  \hat{a}_-  +\beta \sqrt{\kappa_{1}^\mathrm{ext}}\hat{a}_{1}^\mathrm{in} + \alpha \sqrt{\kappa_{2}^\mathrm{ext}} \hat{a}_{2}^\mathrm{in} +  \beta \sqrt{\kappa_{1}^\mathrm{int}}\hat{a}_{1}^\mathrm{int}   + \alpha \sqrt{\kappa_{2}^\mathrm{int}} \hat{a}_{2}^\mathrm{int}  \\
	\dot{\hat{a}}_-	=&  -\left( -i\delta_+ + \frac{\kappa_1}{2} \alpha^2 + \frac{\kappa_2}{2} \beta^2 + i\frac{1}{\alpha\beta}J \right)\hat{a}_-     - \alpha \beta  \Delta \kappa \hat{a}_+  +\alpha \sqrt{\kappa_{1}^\mathrm{ext}} \hat{a}_{1}^\mathrm{in} - \beta \sqrt{\kappa_{2}^\mathrm{ext}} \hat{a}_{2}^\mathrm{in}+\alpha \sqrt{\kappa_{1}^\mathrm{int}} \hat{a}_{1}^\mathrm{int} - \beta \sqrt{\kappa_{2}^\mathrm{int}} \hat{a}_{2}^\mathrm{int} \label{eq:EOMplus-minus},
	\end{align}
	where $\delta_+ \equiv \omega_\text{in} - \omega_+= \Delta + 2J\alpha\beta$ is the detuning from the ``+'' resonance, and $\Delta\kappa \equiv (\kappa_1 - \kappa_2)/2$ (as above). These reduce to those of Ref.~\cite{Yanay2016Quantum} when $|G_1|=|G_2|$, or, equivalently, when $\alpha = \beta = 1/2$. On the other hand, for highly asymmetric coupling $|G_2| \gg |G_1|$ ($|G_1| \gg |G_2|$), we find $\beta \rightarrow 1$ and $\alpha \rightarrow 0$ ($\alpha \rightarrow 1$ and $\beta\rightarrow 0$), such that the eigenmodes effectively decouple from one of the inputs, a behavior of central importance in suppressing measurement backaction.

	\subsection{Measurement Backaction with Quadratic Dispersive Coupling}\label{app:A2}

	
	\textbf{Optical Susceptibilities:}
	In the frequency domain, the equations of motion in the original sub-cavity basis $\hat{a}_j$ (Eq.~\ref{eq:EOM}) can be written
	\begin{align}
	\begin{pmatrix}
	\hat{a}_1(\omega)\\
	\hat{a}_2(\omega)
	\end{pmatrix}	
	=
	&(   -i\omega \boldsymbol{I} +i \boldsymbol{\tilde{H}} )^{-1}   
	\left[ \boldsymbol{\tilde{\kappa}^\text{ext}}
	\begin{pmatrix}
	\hat{a}_{1}^\mathrm{in}(\omega)\\
	\hat{a}_{2}^\mathrm{in}(\omega)
	\end{pmatrix}
	+
	\boldsymbol{\tilde{\kappa}}^\mathrm{int}
	\begin{pmatrix}
	\delta\hat{a}_{1}^\mathrm{int}(\omega)\\
	\delta\hat{a}_{2}^\mathrm{int}(\omega)
	\end{pmatrix}\right],
	\end{align}
	where $ \boldsymbol{I}$ is the identity matrix, and $\hat{a}_j^\mathrm{in}$ comprises any drive and / or fluctuations (including thermal). We can define a susceptibility matrix
	\begin{align}
	\boldsymbol{\tilde{\chi}} (\omega) &= (   -i\omega \boldsymbol{I} +i \boldsymbol{\tilde{H}} )^{-1} \nonumber \\
	&=\frac{1}{\chi_1^{-1}(\omega)\chi_2^{-1}(\omega)+J^2}\begin{pmatrix}
	\chi_2^{-1}(\omega) & iJ \\
	iJ & \chi_1^{-1}(\omega) 
	\end{pmatrix} \label{eq:chi_matrix}  \\
	&\equiv \begin{pmatrix}
	\tilde{\chi}_{11}(\omega) & \tilde{\chi}_{12}(\omega) \\
	\tilde{\chi}_{21}(\omega) & \tilde{\chi}_{22}(\omega) 
	\end{pmatrix} \nonumber
	\end{align}
	where
	\begin{align}
	\chi_j^{-1}(\omega) = -i(\Delta -G_j \Delta x+ \omega) + \kappa_j/2 \label{eq:susceptibilities_general}
	\end{align}
	is the optical susceptibility of the uncoupled ($J=0$) sub-cavity $j$, and $\tilde{\chi}_{ij}$ is shorthand notation for the matrix elements. At the quadratic point $\Delta x = \Delta x_+$, the sub-cavity susceptibilities become
	\begin{align}
	\chi_1^{-1}(\omega) &= -i\left(\delta_+ - J\alpha /\beta + \omega  \right) + \kappa_1/2\\
	\chi_2^{-1}(\omega)&= -i\left(\delta_+ - J \beta /\alpha + \omega  \right) + \kappa_2/2.
	\end{align}

	\textbf{Radiation Force Noise:}
	At the quadratic point $\Delta x_+$, the optical force operator $\hat{F}_\mathrm{opt} = -d\hat{H}_\mathrm{opt}/dx$ reduces to
	\begin{align}
	\hat{F}_\mathrm{opt}(t) &= -\hbar \left( G_1 \hat{a}_1^\dagger(t) \hat{a}_1(t) + G_2 \hat{a}_2^\dagger(t) \hat{a}_2(t) \right).
	\end{align}
	The corresponding power spectral density can be written as \cite{Clerk2010Introduction}\footnote{We use the Fourier transform convention $X(\omega) = \int_{-\infty}^\infty e^{+i\omega t} X(t) dt$ and $X(t) = \frac{1}{2\pi}\int_{-\infty}^\infty e^{-i\omega t} X(\omega) d\omega$, defining $X^\dagger(\omega) \equiv \int_{-\infty}^\infty e^{+i\omega t} X^\dagger(t) dt$; thus, $X^\dagger(\omega)$ refers to the Fourier transform of the time domain variable $X^\dagger(t)$. Note also that $X^\dagger(\omega) = [X(-\omega)]^\dagger$ in this convention.}
	\begin{align}\label{Eq:ForceNoise} 
	S_{FF} (\omega) = \frac{1}{2\pi} \int_{-\infty}^\infty \left \langle  \delta \hat{F}^\dagger_\mathrm{opt}(\omega) \delta \hat{F}_\mathrm{opt}(\omega') \right \rangle   d\omega',
	\end{align} 
	where $\delta \hat{F}_\mathrm{opt}(\omega)$ is the Fourier transform of the fluctuations $\delta \hat{F}_\mathrm{opt}(t) \equiv \hat{F}_\mathrm{opt} (t) -  \langle F_\mathrm{opt} \rangle $ about the mean (i.e. time-averaged) $\langle F_\mathrm{opt} \rangle$. Similarly, expressing $\hat{a}_j(t) = \bar{a}_j + \delta\hat{a}_j(t)$ in terms of fluctuations $\delta\hat{a}_j(t)$ about the mean $\bar{a}_j$, we can linearize the optical force operator for small $\delta\hat{a}_j$, yielding
	\begin{align}\label{Eq:Force}
	\delta \hat{F}_\mathrm{opt}(t) \approx  - \hbar \left(G_1\bar{a}_1^* \delta\hat{a}_1(t)  +  G_2 \bar{a}_2^* \delta\hat{a}_2(t) \right) + \mathrm{c.c.},
	\end{align}
	such that (similar to Ref.~\cite{Yanay2016Quantum}) its Fourier transform becomes
	\begin{align}\label{Eq:ForceSimple}
	\delta \hat{F}_\mathrm{opt}(\omega)  = \sum_{j = 1,2}  &\left( \sqrt{\kappa_{j}^\mathrm{ext}}  \mathcal{A}_j(\omega) \delta\hat{a}_{ j}^\mathrm{in} (\omega) + \sqrt{\kappa_{j}^\mathrm{ext}}\mathcal{A}_j^*(-\omega)\delta\hat{a}_{j}^{\mathrm{in}, \dagger}(\omega)  +    \sqrt{\kappa_{j}^\mathrm{int}}  \mathcal{A}_j(\omega) \delta\hat{a}_{j}^\mathrm{int} (\omega) + \sqrt{\kappa_{l}^\mathrm{int}}\mathcal{A}_j^*(-\omega)\delta\hat{a}_{j}^{\mathrm{int},\dagger}(\omega) \right)
	\end{align}
	with coefficients
	\begin{align}
	\mathcal{A}_1 &=  -\hbar  \left( G_1 \tilde{\chi}_{11}(\omega)  \bar{a}_1^*  + G_2 \tilde{\chi}_{21}(\omega) \bar{a}_2^*   \right) \\
	\mathcal{A}_2 &=-\hbar  \left( G_1 \tilde{\chi}_{12}(\omega)   \bar{a}_1^*  + G_2 \tilde{\chi}_{22}(\omega)  \bar{a}_2^*   \right),
	\end{align}
	where $\tilde{\chi}_{nm}$ are the matrix elements of the optical susceptibility $\boldsymbol{\tilde{\chi}}$ (Eq.~\ref{eq:chi_matrix}). Assuming the usual input noise correlators \cite{Clerk2010Introduction}
	\begin{align}
	\langle \delta\hat{a}_{i}^\mathrm{in} (\omega) \delta\hat{a}_{j}^\mathrm{in} (\omega')\rangle &=\langle \delta\hat{a}_{i}^{\mathrm{in},\dagger} (\omega) \delta\hat{a}_{j}^{\mathrm{in},\dagger} (\omega')\rangle  = 0
	\\ 
	\langle \delta\hat{a}_{i}^\mathrm{in} (\omega) \delta\hat{a}_{j}^{\mathrm{in},\dagger} (\omega')\rangle &= 2\pi\delta_{ij} \delta(\omega+\omega')(\bar{n}_{j}^\mathrm{in}+1)
	\\ 
	\langle \delta\hat{a}_{i}^{\mathrm{in},\dagger} (\omega) \delta\hat{a}_{j}^\mathrm{in} (\omega')\rangle &=2\pi\delta_{ij}\delta(\omega+\omega')(\bar{n}_{j}^\mathrm{in}),
	\end{align}
	where $\bar{n}_{j}^\mathrm{in}$ is the mean thermal occupation of the input port bath; the same relations hold for the noise operators of the loss channels by changing subscript  ``in'' $\rightarrow$ ``int'' throughout (though these ports are considered purely thermal). If the baths are in their ground state ($\bar{n}_{j}^\mathrm{in} = \bar{n}_{j}^\mathrm{int} \approx 0$), the force noise simplifies to
	\begin{align}
	S_{FF}(\omega) &= \kappa_1 |\mathcal{A}_1(\omega)|^2 + \kappa_2   |\mathcal{A}_2(\omega)|^2.\label{eq:force-noise-0K}
	\end{align} 
	In this work, we are particularly interested in the force noise when one eigenmode, having purely quadratic dispersive coupling, is driven by a single port. For example, if $\Delta x=\Delta x_+$ and the system is driven by $\bar{a}_1^\mathrm{in}$ (with $\bar{a}_{2}^\mathrm{in}=0$), the steady-state amplitudes simplify to
	\begin{align}
	\bar{a}_1 &= \tilde{\chi}_{11}(0)\sqrt{\kappa_{1}^\mathrm{ext}} \bar{a}_{1}^\mathrm{in}\label{eq:MeanCavityField1}\\
	\bar{a}_2 &= \tilde{\chi}_{21}(0)\sqrt{\kappa_{1}^\mathrm{ext}} \bar{a}_{1}^\mathrm{in}.\label{eq:MeanCavityField2}
	\end{align}
	Together with Eq.~\ref{eq:eigenmodes}, the ratio $\bar{a}_1/\bar{a}_2$ now allows us to express $\bar{a}_j$ in terms of $\bar{a}_+$ as
	\begin{align}\label{eq:mean_field_a+}
	\bar{a}_1 &=\frac{1}{\beta + i\alpha J \chi_2(0)} \bar{a}_+  \\
	\bar{a}_2 &= \frac{iJ\chi_2(0)}{\beta + i\alpha J \chi_2(0)} \bar{a}_+,
	\end{align}
	which, finally, allows us to write the force noise (Eq.~\ref{eq:force-noise-0K}) in terms of $|\bar{a}_+|^2$:
	\begin{align}\label{eq:BackactionForceNoiseFull-Appendix}
	S_{FF} (\omega) =  \hbar^2 |\bar{a}_+|^2 \frac{1}{\left| \beta + i\alpha J \chi_2 (0) \right|^2} & \left( \kappa_1  \left| G_1\tilde{\chi}_{11}(\omega) - i G_2 J\chi_2^*(0)\tilde{\chi}_{21}(\omega)  \right|^2+ \kappa_2  \left| G_1 \tilde{\chi}_{12}(\omega) - i G_2 J\chi^*_2(0)\tilde{\chi}_{22}(\omega)\right|^2 \right).
	\end{align}
	In the large-gap limit, where $4J\alpha\beta \gg \kappa_1$, $\kappa_2$, $|\omega|$, $|\delta_+|$, such that only the ``+'' mode is involved, at zero detuning this simplifies to
	\begin{align}
	S_{FF}(\omega) =  \hbar^2 |\bar{a}_+|^2 & \frac{ G_1^2 G_2}{4J^2(G_2 -G_1)^2} \frac{4\omega^2 G_2 \kappa_- + (G_2 - G_1) \kappa_+^2 \kappa_2}{\omega^2 + \kappa_+^2/4}.
	\end{align}
	If we instead consider a single-port cavity ($\kappa_2  = 0$ and $\kappa_1 = \kappa_{1}^\mathrm{ext}$), Eq.~\ref{eq:BackactionForceNoiseFull-Appendix} becomes
	\begin{align}\label{eq:BackactionForceNoise_Large-Gap-Appendix}
	S_{FF} (\omega) =  \hbar^2 |\bar{a}_+|^2 & \frac{  \kappa_{1}^\mathrm{ext}}{\left| \beta + i\alpha J \chi_2 (0) \right|^2} \left| \frac{  G_1\chi_2^{-1}(\omega) + G_2 J^2\chi_2^*(0) }{\chi_1^{-1}(\omega)\chi_2^{-1}(\omega)+J^2}  \right|^2,
	\end{align}
	which reduces to
	\begin{align}
	S_{FF} (\omega) = \frac{-\hbar^2 |\bar{a}_+|^2 \kappa_+  G_1^3G_2}{J^2(G_2 - G_1)^2}\frac{(2\delta_+ + \omega)^2}{(\delta_+ +\omega)^2 + \kappa_+^2/4}
	\end{align}
	in the large-gap limit. This expression agrees with the force noise calculated from the simple assumption of dissipative coupling \cite{Elste2009Quantum} when $\tilde{B}$ is given by Eq.~\ref{eq:DissipativeCouplingB}. 
	
	\subsection{Force Noise Reduction}\label{app:Force_Noise_Reduction}
	
	To see how asymmetric sub-cavity LDC can reduce quantum radiation force noise (QRFN), we now consider a membrane-cavity style system. In the large-gap limit, the QRFN (Eq.~\ref{eq:BackactionForceNoise_Large-Gap-Appendix}) is
	\begin{align}
	S_{FF}(\omega) 	&= \frac{\hbar^{2}|\bar{a}_{+}|^{2}\omega_{0}^{2}}{c^{2}|t_{m}|^{2}}  \frac{ L_{2}}{L^{2}}  \frac{4L_{1}\kappa_{-}\omega^{2}+L\kappa_{+}^{2}\kappa_{2}}{\omega^{2}+\kappa_{+}^{2}/4}\label{eq:force-noise-large-gap-membrane-Appendix}, 
	\end{align}
	with decay rates
	\begin{align}
	\kappa_+  &= \frac{L_1}{L} \kappa_1  + \frac{L_2}{L} \kappa_2 \\
	\kappa_- &= \frac{L_2}{L} \kappa_1   + \frac{L_1}{L} \kappa_2  \label{eq:kappa-minus-app},
	\end{align}
	at the quadratic point $\Delta x_+$. At the mechanical frequency $\Omega_m$, this be rewritten
	\begin{align}
	S_{FF}(\Omega_m) 	&= 4 \frac{\hbar^{2}|\bar{a}_{+}|^{2}\omega_{0}^{2}}{c^{2}|t_{m}|^{2}}  \frac{ L_{1} L_{2}}{L^{2}} \kappa_{-} \frac{L \kappa_{2}/(L_1\kappa_-)  + 4\Omega_m^{2}/\kappa_+^2}{1 + 4\Omega_m^{2}/\kappa_+^2}.
	\end{align}
	This expression can be minimized with respect to $L_1$ for fixed cavity length, by substituting the expressions for the decay rates (Eqs.~\ref{eq:kappa0} and \ref{eq:kappa_-_MIC}), yielding  
	\begin{align}
	S_{FF}^\mathrm{min}(\Omega_m) 
	&=2 \frac{\hbar^{2}|\bar{a}_{+}|^{2}\omega_{0}^{2}}{cL|t_{m}|^{2}}    (|t_2|^2 + \mathcal{T}_2) \frac{1 +4\mathcal{B}\Omega_m^2/\kappa_+^2}{1+4 \Omega_m^2/\kappa_+^2},
	\end{align}
	where we defined
	\begin{align}
	\mathcal{B} \equiv \frac{|t_1|^2 + \mathcal{T}_1}{|t_1|^2 + \mathcal{T}_1 + |t_2|^2 + \mathcal{T}_2},
	\end{align}
	noting $\mathcal{B} \in [0,1]$ with $\mathcal{B} =1$ corresponding to the ``single-port'' limit, where $|t_1|^2 +\mathcal{T}_1 \gg  |t_2|^2 + \mathcal{T}_2$. This minimum is found a distance
	\begin{align} \label{eq:L1_min_app}
	L_{1, \mathrm{min}} = \mathcal{B} L
	\end{align}
	from the input mirror.
	
	\textbf{``Modified'' resolved-sideband limit}: In the limit $\Omega_m/\kappa_+ \gg \sqrt{L \kappa_{2}/4L_1\kappa_-} $ and $\Omega_m/\kappa_+ \gg 1/2$  -- both of which are succinctly captured by $\sqrt{\mathcal{B}} \Omega_m \gg \kappa_+$ at $L_{1,\mathrm{min}}$ -- the minimal QRFN reduces to
	\begin{align}
	S_{FF}^\mathrm{min, RS}(\Omega_m) =&  2 \frac{\hbar^{2}|\bar{a}_{+}|^{2}\omega_{0}^{2}}{cL|t_{m}|^{2}} \mathcal{B} (|t_2|^2 + \mathcal{T}_2) \\
	\xrightarrow{\text{Single-Port}}&   2 \frac{\hbar^{2}|\bar{a}_{+}|^{2}\omega_{0}^{2}}{cL|t_{m}|^{2}} (|t_2|^2 + \mathcal{T}_2),
	\end{align}
	where the second line is evaluated for small but finite $|t_2|^2+\mathcal{T}_2 \ll |t_1|^2+\mathcal{T}_1$ ($\mathcal{B}\approx 1$). In the latter case, the QRFN is limited simply by the distance between the lossless back mirror and its closest quadratic point (wavelength scale).
	
	\textbf{``Modified'' fast-cavity limit}: In the limit $\Omega_m/\kappa_+ \ll \sqrt{L \kappa_{2}/4L_1\kappa_-} $   (and $\Omega_m/\kappa_+ \ll 1/2$) -- captured succinctly by $\Omega_m \ll \kappa_+ $  at $L_{1,\mathrm{min}}$ -- the minimal QRFN reduces to
	\begin{align}
	S_{FF}^\mathrm{min, FC}(\Omega_m) 	&=  2 \frac{\hbar^{2}|\bar{a}_{+}|^{2}\omega_{0}^{2}}{cL|t_{m}|^{2}}  (|t_2|^2 + \mathcal{T}_2).
	\end{align}
	which is a factor
	\begin{align}
	\frac{	S_{FF}^\mathrm{min, FC}(\Omega_m) }{	S_{FF}^\mathrm{min, RS}(\Omega_m) }  = \frac{1}{\mathcal{B}}\geq 1
	\end{align}
	larger than that of the resolved-sideband case.

	\subsection{Application: Optical Trapping}\label{app:Optical_Trapping}
	
	The optical spring generated at a quadratic point has the same strength as a free-space trap \cite{Ni2012Enhancement} per watt incident on the membrane, though with significantly less input power due to the cavity enhancement. As discussed below, our scheme realizes this with lower QRFN than can be achieved in free space (or with a MIM system). 
	
	If we write the optical Hamiltonian in the ``+'' mode basis, and expand its frequency $\omega_+$ to second order in mechanical displacement $\hat{x}$, we find
	\begin{align}
	\hat{H}_\mathrm{opt} &=  \hbar\omega_+(x) \hat{a}^\dagger_+ \hat{a}_+ \approx \hbar \left( \omega_+(0) + \frac{1}{2}\frac{\partial^2\omega_+}{\partial x^2} \hat{x}^2  \right) \hat{a}^\dagger_+ \hat{a}_+,
	\end{align}
	which allows us to directly identify the spring constant 
	\begin{align}
	k_\mathrm{opt} &=  \hbar  \frac{\partial^2\omega_+}{\partial x^2} |\bar{a}_+ |^2    =  \frac{4}{|t_m|} \frac{ \hbar \omega_0^2}{cL} |\bar{a}_+ |^2,
	\end{align}
	where we have substituted in the quadratic dispersive coupling $\partial^2_x \omega_+$ from Eq.~\ref{eq:d2omega_+-dx^2_membrane} and considered (for the time being) only the non-dynamical part of the spring (i.e., $\hat{a}_+ \approx \bar{a}_+$). The optical spring can then be expressed as a function of the (mean)  circulating power
	\begin{align}
	\bar{P}_\mathrm{circ} = \hbar \omega_\mathrm{in} \frac{c}{2L}|\bar{a}_+|^2 
	\end{align}
	as
	\begin{align}
	k_\mathrm{opt} &  =  \frac{8}{|t_m|}  \frac{ \omega_\mathrm{in}  P_\mathrm{circ} }{c^2}, 
	\end{align}
	using $\omega_\mathrm{in} \approx \omega_0$, which is the same expression as for a free-space trap with a retro-reflected beam \cite{Ni2012Enhancement}.
	
	When approaching the single-port limit ($|t_2|^2+\mathcal{T}_2\ll |t_1|^2+\mathcal{T}_1$), the minimal force noise at the optimal membrane position is
	\begin{align}\label{eq:trapping-minimum-SFF}
	S_{FF}^\mathrm{min}(\Omega_m) 	&=  \frac{S_{FF}^\mathrm{FS}}{2} \frac{|t_2|^2 + \mathcal{T}_2}{|t_m|^2}  ,
	\end{align}
	regardless of whether the system is sideband-resolved, where
	\begin{align}
	S_{FF}^\mathrm{FS}  \approx 8 \frac{ \hbar \omega_\mathrm{in}  }{c^2} \bar{P}_\mathrm{circ},
	\end{align}
	is the force noise associated with a free-space trap in the limit $|t_m|^2 \ll 1$. $S_{FF}^\mathrm{FS}$ can be obtained noting that, for a highly reflective membrane, the force applied to each side is $F \approx 2P_\mathrm{circ}/c$, yielding a force noise of $S_{FF} = 4 S_{P_\mathrm{circ} P_\mathrm{circ}}/c^2$. For shot noise, the power spectral density is $S_{P_\mathrm{circ} P_\mathrm{circ}} = \hbar \omega_\mathrm{in} \bar{P}_\mathrm{circ}$ and since shot noise is uncorrelated at each side of the membrane, it adds in quadrature, yielding the above equation. Importantly, Eq.~\ref{eq:trapping-minimum-SFF} shows that the force noise can be improved by a factor $2|t_m|^2/(|t_2|^2+\mathcal{T}_2)$; together with the comparative ease of realizing shot-noise-limited input light at lower powers, our approach presents a significant advantage over free-space traps.

	\subsection{Application: QND Measurement} \label{app:A3}
	Here we present the potential advantages our technique provides within the context of quantum nondemolition (QND) phonon number measurements.
	
	\subsubsection{Backaction Rate}
	The rate at which $S_{FF}$ adds or removes a phonon from a mechanical oscillator containing $n$ phonons is \cite{Clerk2010Introduction,Clerk2010Quantum} 
	\begin{align}
	\Gamma_{\mathrm{ba},n} &= \frac{x_\mathrm{zpf}^2}{\hbar^2} [(1+n) S_{FF}(-\Omega_\mathrm{m}) + n S_{FF}(+\Omega_\mathrm{m})].
	\end{align}
	With a resonant drive ($\delta_+= 0$) and in the resolved sideband regime ($\Omega_\mathrm{m} \gg \kappa_+$), this is
	\begin{align}
	\Gamma_{\mathrm{ba},n}  &=  (2n+1) x_\mathrm{zpf}^2 |\bar{a}_+|^2 \kappa_- \frac{G_1^2 G_2^2}{J^2(G_2 -G_1)^2},
	\end{align}
	where we assume $(G_2-G_1)\kappa_2/(4G_2\kappa_-) \ll \Omega_m^2/\kappa_+^2$.\footnote{This holds, e.g., for the membrane-cavity system in the resolved sideband regime ($\Omega_m\gg\kappa_+$) in the limit  $|t_1|^2 + \mathcal{T}_1 \gtrsim |t_2|^2 + \mathcal{T}_2 $ since $(G_2-G_1)\kappa_2/(4G_2\kappa_-)= L^2(|t_2|^2+\mathcal{T}_2)/[4L_1^2(|t_2|^2+\mathcal{T}_2) + 4L_2^2(|t_1|^2 + \mathcal{T}_1)]  \lesssim 1  $.}

	\subsubsection{Measurement Rate}\label{app:Measurement_rate_full}
	
	In a quadratically-coupled optomechanical system, phonons each produce a shift in the resonant frequency of the optical mode \cite{Thompson2008Strong,Jayich2008Dispersive,Bhattacharya2008Optomechanical}. In a resonantly driven cavity, this will produce a shift in the phase of the reflected light. Here we derive the phonon measurement rate for an asymmetric optomechanical system with quadratic dispersive coupling using homodyne detection.

	\subsubsection*{Input-Output Relations}\label{sec:input-output}
	
	To quantify how the phase of the reflected light depends on phonon number, we first relate the mean reflected output field $\bar{a}_1^\mathrm{out}$ to the mean input field $\bar{a}^\mathrm{in}_1$ (assuming we only address sub-cavity 1) using the standard input-output relation  \cite{Clerk2010Introduction}
	\begin{align}
	\bar{a}_{1}^\mathrm{out} &= \bar{a}_{1}^\mathrm{in} -\sqrt{\kappa_{1}^\mathrm{ext} }\bar{a}_1\nonumber \\
	&=\left[1 -  \tilde{\chi}_{11}(0)\kappa_{1}^\mathrm{ext}  \right]\bar{a}_{1}^\mathrm{in} \label{eq:input-ouput-relation},
	\end{align}
	where $\bar{a}_1$ is the mean field in sub-cavity 1; the second line is obtained from Eq.~\ref{eq:MeanCavityField1}. In the large gap limit, where $4J\alpha\beta \gg \kappa_1$, $\kappa_2$, $|\omega|$, $|\delta_+|$, which simplifies the value $\tilde{\chi}_{11}$ given by Eq.~\ref{eq:chi_matrix}, this becomes
	\begin{align}
	\bar{a}_{1}^\mathrm{out}  &= \frac{i\delta_+ - \kappa_+/2 + \beta^2 \nonumber \kappa_{1}^\mathrm{ext}}{i\delta_+ - \kappa_+/2} \bar{a}_{1}^\mathrm{in}\\
	&= A e^{-i\phi}    \bar{a}_{1}^\mathrm{in} ,
	\end{align}
	where, in the second line, we defined the (real) amplitude reflection coefficient
	\begin{align}
	A&\equiv   \sqrt{ \frac{\delta_+^2 +( \beta^2 \kappa_{1}^\mathrm{ext} + \kappa_+/2)^2}{\delta_+^2 + \kappa_+^2/4}} 
	\end{align}
	and the (real) reflected phase
	\begin{align}
	\phi & \equiv \arctan \left\{ \frac{\delta_+ \beta^2 \kappa_{1}^\mathrm{ext}}{\delta^2_+ +\kappa_+^2/4 -\beta^2 \kappa_{1}^\mathrm{ext} \kappa_+/2}  \right\}.
	\end{align}
	For a resonant drive frequency $\omega_\textrm{in}$ (detuning $\delta_+=0$), this phase changes as
	\begin{align}
	\left. \frac{d \phi}{d\omega_+}  \right|_{\delta_+ = 0}= - \left. \frac{d\phi}{d\delta_+} \right|_{\delta_+ =0}=  \frac{-\beta^2\kappa_{1}^\mathrm{ext}}{\kappa_+^2/4 - \beta^2 \kappa_{1}^\mathrm{ext}\kappa_+/2 } \label{eq:dphi_domega_+},
	\end{align}
	We can also relate the mean field in the ``+'' mode to the input field using Eqs.~\ref{eq:eigenmodes}, \ref{eq:MeanCavityField1} and \ref{eq:MeanCavityField2}:
	\begin{align}\label{eq:a_+-a_in}
	\bar{a}_+ = \frac{i\alpha J + \beta \chi_2^{-1}(0)}{J^2+\chi_1^{-1}(0) \chi_2^{-1}(0)} \sqrt{\kappa_{1}^\mathrm{ext}} \bar{a}_{1}^\mathrm{in},
	\end{align}
	so that the output field (Eq.~\ref{eq:input-ouput-relation}) can be written in terms of  ``+'' mode as
	\begin{align}\label{eq:a_out-a_+}
	\bar{a}_{1}^\mathrm{out} =\frac{J^2+\chi_1^{-1}(0) \chi_2^{-1}(0)  - \kappa_{1}^\mathrm{ext} \chi_2^{-1}(0)}{i\alpha J + \beta \chi_2^{-1}(0)} \frac{\bar{a}_+}{\sqrt{\kappa_{1}^\mathrm{ext}}}.
	\end{align}
	In the large-gap limit, this reduces to
	\begin{align}
	\bar{a}_{1}^\mathrm{out} = \left ( \kappa_+/2 - \beta^2\kappa_{1}^\mathrm{ext} \right) \frac{\bar{a}_+}{\sqrt{\beta^2 \kappa_{1}^\mathrm{ext}}} \label{eq:a_out-a_+_largegap}.
	\end{align}

	\subsubsection*{Measurement Rate with Homodyne Detection}
	To derive the measurement rate, we consider an ideal homodyne measurement (all the photons are collected, the laser is shot-noise limited, and the detection scheme has identical arms) where a local-oscillator (``LO''; mean optical power $\bar{P}^{\mathrm{LO}}$) is combined at a beamsplitter with a measurement beam leaving the cavity (mean power $\bar{P}_{1}^\mathrm{out}$) with the aim of resolving its fluctuating phase $\delta \phi(t)$. Specifically, suppose one beamsplitter input has classical LO field $E^\mathrm{LO} = \sqrt{2\bar{P}^\mathrm{LO}}\cos(\omega_\mathrm{in} t + \Delta \phi )$ with some fixed phase $\Delta \phi$, and the other has the signal field $E_{1}^\mathrm{out} = \sqrt{2\bar{P}_{1}^\mathrm{out}}\cos(\omega_\mathrm{in} t + \delta \phi(t))$, both oscillating at frequency $\omega_\mathrm{in}$. The homodyne signal, obtained by subtracting the photocurrents measured at each of the beamsplitter outputs, is then
	\begin{align}
	\delta I(t) = 2A \sqrt{ \bar{P}^\mathrm{LO} \bar{P}_{1}^\mathrm{out}}\sin(\delta \phi (t) - \Delta \phi),
	\end{align}
	where $A$ is a gain relating optical power to photocurrent.
	Tuning the LO phase to maximize the sensitivity to phase (e.g. $\Delta \phi = 0$) and Taylor expanding $\sin$ yields
	\begin{align}\label{eq:deltaI}
	\delta \phi(t)  \approx \left(2A \sqrt{\bar{P}_{\mathrm{out},1} \bar{P}_{\mathrm{LO}}}\right)^{-1}\delta I(t).
	\end{align}
	At the same time, the combined shot noise power spectral density $S^\textrm{sn}_{II}$ from the two (subtracted) photocurrents provides a noise floor
	\begin{align}\label{eq:Sphiphi1}
	S_{\phi \phi} &=  (4A^2\bar{P}_{1}^\mathrm{out} \bar{P}^{\mathrm{LO}} )^{-1} S_{II}^\mathrm{sn},
	\end{align}
	and, for large LO ($\bar{P}^{\mathrm{LO}} \gg \bar{P}_{1}^\mathrm{out} $), 
	\begin{align}\label{eq:SII1}
	S_{II}^\mathrm{sn} \approx A^2 \hbar\omega_{\mathrm{in}}\bar{P}^{\mathrm{LO}}.
	\end{align}
	Furthermore, since a frequency fluctuation $\delta \omega_+$ of the cavity produces phase $\delta \phi \approx ( d \phi/d\omega_+)\delta \omega_+ $ ($d\phi/d\omega_+$ derived as above), the frequency noise floor
	\begin{align}
	S_{\omega_+ \omega_+}&= S_{\phi\phi}\left( \frac{d\phi }{d\omega_+} \right)^{-2} = \frac{1}{4  |\bar{a}_{1}^\mathrm{out}|^2} \left( \frac{d\phi }{d\omega_+} \right)^{-2},
	\end{align}
	where, in the last step, we substituted in Eqs.~\ref{eq:Sphiphi1}-\ref{eq:SII1} and expressed the mean signal power $\bar{P}_{1}^\mathrm{out} = \hbar \omega_\mathrm{in} |\bar{a}_{1}^\mathrm{out}|^2$ in term of the reflected photon rate $|\bar{a}_{1}^\mathrm{out}|^2$, in accordance with Ref.~\cite{Clerk2010Introduction}. 
	
	To convert this result to the measurement time $t_\mathrm{meas}$ required to resolve a frequency shift $\Delta \omega_+$, we compare the noise floor to the frequency shift, requiring
	\begin{align}
	t_\mathrm{meas} \geq  \frac{S_{\omega_+\omega_+}}{ (\Delta\omega_+)^2},
	\end{align}
	in order to resolve the frequency shift with unity signal-to-noise ratio. For a single phonon jump, the induced frequency shift is 
	\begin{align}
	\Delta \omega_+ = \frac{d^2\omega_+}{dx^2}x_\mathrm{zpf}^2,
	\end{align}
	yielding a measurement rate 
	\begin{align}
	\Gamma_\mathrm{meas} \equiv \frac{1}{t_\mathrm{meas}} &= 4 |\bar{a}_{1}^\mathrm{out}|^2 \left( \frac{d\phi}{d\omega_+}  \frac{d^2\omega_+}{dx^2}x_\mathrm{zpf}^2 \right)^2.
	\end{align}
	In the large gap limit, using Eq.~\ref{eq:d2omega_+-dx^2} for $d^2\omega_+/dx^2$, and Eqs. \ref{eq:dphi_domega_+} and \ref{eq:a_out-a_+_largegap} derived in Sec.~\ref{sec:input-output} to substitute $d\phi/d\omega_+$ and $\bar{a}_{\mathrm{out}, 1}$ respectively, this measurement rate simplifies to
	\begin{align}
	\Gamma_\mathrm{meas} & = \frac{16|\bar{a}_+|^2}{\kappa_+}  \frac{\beta^2\kappa_{1}^\mathrm{ext}}{\kappa_+} \left(   \frac{d^2\omega_+}{dx^2}x_\mathrm{zpf}^2 \right)^2  \\
	&= \frac{64|\bar{a}_+|^2}{\kappa_+}  \frac{\beta^2\kappa_{1}^\mathrm{ext}}{\kappa_+}   \frac{1}{J^2} \frac{(-G_1 G_2)^{3}}{(G_2 - G_1)^2}    x_\mathrm{zpf}^4.
	\end{align}
	The  correction term $\beta^2 \kappa_{1}^\mathrm{in}/\kappa_+$ accounts for the proportion of photons in the ``+'' mode leaving through the input port. 
	
	For the case of a truly single-port cavity ($\kappa_2 = 0$ and $\kappa_1 = \kappa_{1}^\mathrm{ext}$), the measurement rate is
	\begin{align}
	\Gamma_{\mathrm{meas}} &= \frac{16 |\bar{a}_+|^2}{\kappa_+}  \left(   \frac{d^2\omega_+}{dx^2}x_\mathrm{zpf}^2 \right)^2 \nonumber\\
	&= \frac{64 |\bar{a}_+|^2}{\kappa_+} \frac{(-G_1G_2)^3}{J^2(G_2-G_1)^2} x_\mathrm{zpf}^4.
	\end{align}
	Note for a membrane-cavity system, $\kappa_+$, $\partial^2_x\omega_+$, and thus $\Gamma_\mathrm{meas}$, do not depend upon at which quadratic point the membrane is positioned, allowing one to tune the ratio $G_1/G_2$ without affecting the measurement rate.

	\subsubsection{Backaction-Limited Number State Resolution} 
	The measurement rate, which is derived in Sec.~\ref{app:Measurement_rate_full} below, is
	\begin{align}
	\Gamma_\mathrm{meas} &= \frac{64|\bar{a}_+|^2}{\kappa_+}  \frac{\beta^2\kappa_{1}^\mathrm{ext}}{\kappa_+}   \frac{1}{J^2} \frac{(-G_1 G_2)^{3}}{(G_2 - G_1)^2}    x_\mathrm{zpf}^4,
	\end{align}
	yielding a figure of merit
	\begin{align}\label{eq:Meas_to_Back_Fulll}
	\frac{\Gamma_\mathrm{meas}}{\Gamma_{\mathrm{ba},n}} = \frac{64}{2n+1} \frac{g_1 g_2}{\kappa_- \kappa_+} \frac{\beta^2\kappa_{1}^\mathrm{ext}}{\kappa_+},
	\end{align}
	describing how well a phonon number state $n$ can be measured before backaction causes a jump, where we define the usual optomechanical coupling rate $g_i \equiv x_\mathrm{zpf} |G_i|$. In the single-port cavity limit ($\kappa_2=0$), this ratio becomes
	\begin{align}
	\frac{\Gamma_\mathrm{meas}}{\Gamma_{\mathrm{ba},n}} &= \frac{64}{2n+1} \frac{(g_2 + g_1)^2 }{\kappa_1^2}  = \frac{64}{2n+1} \frac{g_2^2 }{\kappa_+^2} .
	\end{align}
	The advantage of our approach becomes clearer if we consider a membrane-cavity system in which the membrane is a distance $L_1$ from the first mirror and $L_2$ from the second (total length $L=L_1+L_2$). If the sub-cavity crossing frequency $\omega_0 = N_1 \pi c/L_1 = N_2 \pi c/L_2$ where $N_i = 2L_i/\lambda$ is the (integer) mode index of each sub-cavity at wavelength $\lambda$, and the coupling rates  $G_1 = -\omega_0/L_1$ and $G_2 = +\omega_0/L_2$, the ratio becomes
	\begin{align}
	\frac{\Gamma_{\mathrm{meas}}}{\Gamma_{\mathrm{ba},n}} &= \frac{64}{2n+1} \left(\frac{  \omega_0 x_\mathrm{zpf}}{L_2 \kappa_0}\right)^2 \nonumber \\
	&=\frac{64}{2n+1}\left(\frac{g_{\text{MIM}}}{\kappa_{0}}\right)^{2}\left(\frac{L}{2L_{2}}\right)^{2}
	\end{align}
	with $\kappa_0$ being the empty cavity power decay rate (Eq.~\ref{eq:kappa0}). The second line is expressed in terms of the half-cavity single-photon optomechanical coupling rate for the MIM system $g_\text{MIM} \equiv 2x_\textrm{zpf}\omega_0/L$ to facilitate a direct comparison: by reducing $L_2$ (moving the membrane toward the back mirror), the usual strong coupling requirement $g_\textrm{MIM}/\kappa_0 > 1$ for measuring a state before QRFN destroys it \cite{Yanay2016Quantum} is relaxed by a factor $L/2L_2$, which can be as large as $\sim L/\lambda$ at the quadratic point nearest the back mirror.
	
	For fixed mirror losses (round trip power loss $\mathcal{T}_j=\mathcal{T}$, say) we can improve the fidelity by a less dramatic factor, and there is some advantage to be gained by bringing $|t_1|^2$ closer to $\mathcal{T}$. To calculate the optimal $L_{1, \mathrm{min}}$ and $t_{1,\mathrm{opt}}$, we first substitute the expression for the decay rates (Eqs.~\ref{eq:kappa0} and \ref{eq:kappa_-_MIC}), and optomechanical coupling $g_i =  x_\mathrm{zpf}\omega_0/L_j$ in the ratio Eq.~\ref{eq:Meas_to_Back_Fulll}; the optimal membrane position is then
	\begin{align}\label{eq:optimized_x}
	L_{1, \mathrm{min}} = \frac{|t_1|^2 + \mathcal{T}_1 }{|t_1|^2 + \mathcal{T}_1 + |t_2|^2 + \mathcal{T}_2 } L
	\end{align}
	where the force noise is minimal and the ratio
	\begin{align}\label{eq:optimized_ratio}
	\left. \frac{\Gamma_\mathrm{meas}}{\Gamma_{\mathrm{ba},n}} \right|_{L_{1,\mathrm{min}}}= \frac{256}{2n+1} \frac{\omega_0^2x_\mathrm{zpf}^2}{c^2} \frac{|t_1|^2}{(|t_1|^2 + \mathcal{T}_1)(|t_2|^2 + \mathcal{T}_2)(|t_1|^2 + \mathcal{T}_1 +|t_2|^2 +  \mathcal{T}_2 )},
	\end{align}
	is maximal; $L_{1, \mathrm{min}}$ is the same as before  (Eq.~\ref{eq:L1_min_app}) since the measurement rate is independent of the membrane position within the cavity. We can also compare with the same ratio for the membrane-in-the-middle,
	\begin{align}\label{eq:optimized_ratio_MIM}
	\left. \frac{\Gamma_\mathrm{meas}}{\Gamma_{\mathrm{ba},n}} \right|_{L/2}=  \frac{1024}{2n+1}  \frac{ \omega_0^2   x_\mathrm{zpf}^2 }{c^2  }      \frac{|t_1|^2 }{(|t_1|^2 + \mathcal{T}_1 + |t_2|^2 + \mathcal{T}_2)^3 }  ,
	\end{align}
	finding our approach yields an improvement
	\begin{align}\label{eq:resolving-improvement_MIM}
	\left. \frac{\Gamma_\mathrm{meas}}{\Gamma_{\mathrm{ba},n}} \right|_{L_{1, \text{min}}}	\left/ \frac{\Gamma_\mathrm{meas}}{\Gamma_{\mathrm{ba},n}} \right|_{L/2} =   \frac{1}{4}   \frac{(|t_1|^2 + \mathcal{T}_1 + |t_2|^2 + \mathcal{T}_2)^2 }{(|t_1|^2 + \mathcal{T}_1)(|t_2|^2 + \mathcal{T}_2)} \rightarrow \frac{1}{4}   \frac{|t_1|^2}{|t_2|^2 + \mathcal{T}_2}.
	\end{align}
	As expected, this is the same improvement as we found for the force noise.
	
	With the membrane at the optimal $L_{1,\mathrm{min}}$, we can also calculate the the optimal input mirror transmission
	\begin{align}\label{eq:optimized_T1-Appendix}
	|t_{1}|^2_\textrm{opt} = \sqrt{\mathcal{T}_1 (|t_2|^2 + \mathcal{T}_1 + \mathcal{T}_2 )},
	\end{align}
	that maximizes the ratio, yielding
	\begin{align}\label{eq:ratio_improvement}
	\left. \frac{\Gamma_\mathrm{meas}}{\Gamma_{\mathrm{ba},n}} \right|_{L_{1,\text{min}}, |t_{1}|^2_\textrm{opt}}= \frac{256}{2n+1} \frac{\omega_0^2x_\mathrm{zpf}^2}{c^2} \frac{|t_2|^2+ 2 \mathcal{T}_1 + \mathcal{T}_2-2\sqrt{\mathcal{T}_1(|t_2|^2+\mathcal{T}_1+\mathcal{T}_2)}}{(|t_2|^2+\mathcal{T}_2)^3}.
	\end{align}
	This optimized QND ``fidelity'' is a factor
	\begin{align}\label{eq:resolving-improvement}
	\left. \frac{\Gamma_\mathrm{meas}}{\Gamma_{\mathrm{ba},n}} \right|_{L_{1, \text{min}}, |t_{1}|^2_\textrm{opt}}	\left/ \frac{\Gamma_\mathrm{meas}}{\Gamma_{\mathrm{ba},n}} \right|_{L_{1,\text{min}}, \text{matched}} =2+6\frac{\mathcal{T}}{|t_2|^2}-4\sqrt{\frac{\mathcal{T}}{|t_2|^2}\left(1+2\frac{\mathcal{T}}{|t_2|^2}\right)} 
	\end{align}
	larger than that of a traditional matched cavity ($|t_1|$ set equal to $|t_2|$). Equation \ref{eq:resolving-improvement} can be viewed as a prefactor that generalizes the existing ``standard quantum limit'' \cite{Miao2009Standard} to the case of asymmetric systems. The largest gains occur in systems maximing the back-mirror reflectivity, which is just a statement that one needs to open the input mirror enough to get a reasonable fraction of the cavity light out, and up to 3 dB improvement can also be achieved with transmission-dominated mirrors ($\mathcal{T}\ll|t_j|^2$), though this assumes we ignore the information in the light leaving the back mirror. 
	
	For a fair comparison with MIM systems, it is also possible to calculate the optimal transmission
	\begin{align}
	|t_1|^2_{\mathrm{MIM}} =  (|t_2|^2 + \mathcal{T}_1 + \mathcal{T}_2)/2
	\end{align}
	maximizing the measurement rate to backaction rate ratio for the membrane-in-the middle (Eq.~\ref{eq:optimized_ratio_MIM}), yielding a ratio
	\begin{align}\label{eq:optimized_ratio_MIM_t1_opt}
	\left. \frac{\Gamma_\mathrm{meas}}{\Gamma_{\mathrm{ba},n}} \right|_{L/2, |t_1|^2_{\mathrm{MIM}}}=  \frac{4096}{27(2n+1)}  \frac{ \omega_0^2   x_\mathrm{zpf}^2 }{c^2  }      \frac{1 }{(|t_2|^2 + \mathcal{T}_1 +  \mathcal{T}_2)^2 }  ,
	\end{align}
	and which is a factor
	\begin{align}\label{eq:resolving-improvement_fair_comparison}
	\frac{ \left. \frac{\Gamma_\mathrm{meas}}{\Gamma_{\mathrm{ba},n}} \right|_{L_{1, \text{min}}, |t_{1}|^2_\textrm{opt}}}{	\left. \frac{\Gamma_\mathrm{meas}}{\Gamma_{\mathrm{ba},n}} \right|_{L/2, |t_1|^2_{\mathrm{MIM}}}} = \frac{27}{16}   \frac{(|t_2|^2 + \mathcal{T}_1 + \mathcal{T}_2)^2(|t_2|^2+2\mathcal{T}_1+\mathcal{T}_2-2\sqrt{\mathcal{T}_1(|t_2|^2+\mathcal{T}_1+\mathcal{T}_2)})}{(|t_2|^2+\mathcal{T}_2)^3} 
	\end{align}
	lower than at the optimal membrane position.

	\section{Hopping Rate for a Membrane-in-Cavity}\label{app:HoppingRate}
	In this section, we derive the classical equations of motion for the electric fields for a cavity with a membrane inside it, extending the formalism of Refs.~\cite{Lawrence1999Dynamic, Wilson2011Cavity, Rakhmanov2000Dynamics}. This allows for a simple derivation of EOMs and direct access to the hopping rate $J$ between two sub-cavities separated by a partial reflector.
	\begin{figure}
		\centering
		\includegraphics[width=0.4\columnwidth]{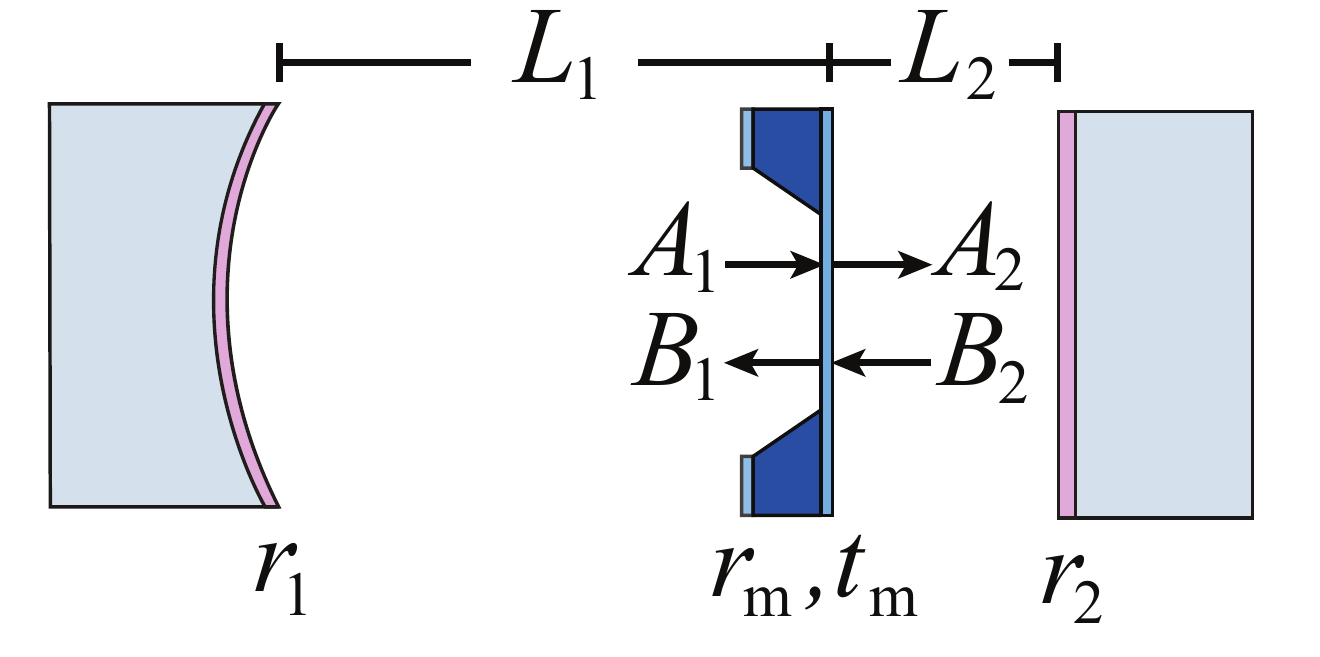}
		\caption{A Fabry-P\'{e}rot cavity comprising a left-hand and right-hand mirror with field reflection coefficients $r_1$ and $r_2$, split into two sub-cavities of length $L_1$ and $L_2$ by a membrane with field reflection (transmission) coefficients $r_m$ ($t_m$). $A_j$ indicate the right moving fields and $B_j$ the left moving fields just outside the membrane surfaces. }
		\label{Fig4}
	\end{figure}

	Consider the Fabry-Perot cavity in Fig.~\ref{Fig4}, comprising two end mirrors of field reflection (transmission) coefficients $r_j$ ($t_j$) partitioned by a third mirror (membrane) having field reflection (transmission) coefficient $r_m$ ($t_m$), such that the sub-cavity lengths $L_j$ sum to total length $L = L_1 + L_2$. The right-moving field amplitude $A_2$ just to the right of the membrane at a time $t$ is related to the left and right incoming amplitudes $A_1$ and $B_2$ as
	\begin{align}\label{eq:A2}
	A_2(t) = r_m B_2(t) + t_m A_1(t),
	\end{align}
	Following this wave a round-trip time  $\tau_{2}= 2L_2/c$ later in sub-cavity 2 ($c$ is the speed of light), we find a returning field
	\begin{align}\label{eq:B2-A2}
	B_2(t+\tau_2) = r_2 e^{i\omega_\text{in} \tau_2}A_2(t),
	\end{align}
	in a frame rotating at the laser frequency $\omega_\mathrm{in}$. Substituting Eq.~\ref{eq:A2} in Eq.~\ref{eq:B2-A2}, we obtain
	\begin{align}
	B_2(t+\tau_2) = r_m r_2 e^{i\omega_\text{in}\tau_2} B_2(t) + t_m r_2 e^{i\omega_\text{in} \tau_2} A_1(t)
	\end{align}
	For small $\tau_{2}$\footnote{We assume the round trip time is small compared to the other relevant time-scales of the system, such that $A_j$ and $B_j$ are slowly varying. Specifically, $1/\tau_2 \gg \kappa_1, \kappa_2, \Omega_\mathrm{m}, J$; the first two conditions arise from $|t_1|, |t_2| \ll 1$ and the last one arises from $|t_m|\ll \sqrt{L_1/L_2}$.}, we can approximate the time derivative of $B_2$ as
	\begin{align}
	\dot{B}_2(t) &\approx \frac{B_{2}(t + \tau_{2}) -  B_2(t)}{\tau_2} \nonumber \\
	& = - \frac{1-r_2 r_{m} e^{i\omega_\text{in}\tau_2  } }{\tau_2}B_{2}(t)+\frac{t_{m} r_2  e^{i\omega_\text{in}\tau_2}}{\tau_2} A_{1}(t).
	\end{align}
	Following the same method for the left sub-cavity field, we find
	\begin{align}
	\dot{A}_1(t)\approx - \frac{(1-r_m r_1 e^{i\omega_\text{in} \tau_1})}{\tau_1} A_1(t) + \frac{t_m r_1 e^{i \omega_\text{in} \tau_1}}{\tau_1} B_2(t).
	\end{align}
	As a matter of convention, we then rescale the fields as
	\begin{align}
	\alpha_1(t) &= - \sqrt{L_1} A_1(t)\\
	\alpha_2(t) &= + \sqrt{L_2} B_2(t)
	\end{align}
	such that $|\alpha_j|^2$ represents the number of photons in sub-cavity $j$. In a matrix form, these coupled equations then become
	\begin{align}
	\begin{pmatrix}
	\dot{\alpha}_1(t) \\
	\dot{\alpha}_2(t) 
	\end{pmatrix}
	&=
	\begin{pmatrix}
	- \frac{c}{2L_1}(1 - r_1 r_{m} e^{i\omega_\text{in}\tau_1 }) & \frac{-r_1e^{i\omega_\text{in}\tau_1}ct_m}{2\sqrt{L_1L_2}}\\
	\frac{ -r_2 e^{i\omega_\text{in}\tau_2}  ct_m}{2\sqrt{L_1L_2}} & - \frac{c}{2L_2}(1 - r_2 r_{m} e^{i\omega_\text{in}\tau_2  })
	\end{pmatrix}
	\begin{pmatrix}
	\alpha_1(t) \\
	\alpha_2(t) 
	\end{pmatrix},
	\end{align}
	in agreement with the more careful derivation of Ref.~\cite{Lang1986Local}.
	
	To recover the equations obtained from input-output theory, we make additional approximations. First, for simplicity, we now make a common unitarity preserving choice for the phase on all the coefficients, such that $r_m = -|r_m|$ and $t_m = i|t_m|$. We furthermore assume that the end mirrors have high reflectivity such that $r_j \approx -1 + |t_j|^2/2$, and write the laser frequency as
	\begin{align}
	\omega_\text{in} = \Delta + \omega_j(\Delta x) - G_j \Delta x,
	\end{align}
	where $\Delta \equiv \omega_\text{in} - \omega_0$ is the detuning from the crossing frequency (defining $\Delta x\equiv 0$), $\omega_j(\Delta x)$ is the resonant frequency of sub-cavity $j$, and $G_j = \partial_x\omega_j$ is the usual linear dispersive coupling. Note that, by definition, the sub cavities have an integer $N_j =2L/\lambda_j$ half wavelengths, such that $\omega_j(\Delta x) \tau_j = 2 \pi N_j $. Again that the detuning is much smaller than the round-trip rate of each sub-cavity  ($|\Delta| \sim \kappa_j \ll 1/\tau_j$), that the membrane is displaced by much less than a half wavelength ($\Delta x \ll \lambda/2$, or, equivalently, $|G_j \Delta x | \ll 1/\tau_j$), and keeping only the terms to leading order in $|t_m|$, $|t_1|^2$, $|t_2|^2$, and $\Delta$,
	\begin{align}
	\begin{pmatrix}
	\dot{\alpha}_1(t) \\
	\dot{\alpha}_2(t) 
	\end{pmatrix}
	&=
	\begin{pmatrix}
	i(\Delta -G_1 \Delta x) - \kappa_1/2 & i \frac{c|t_m|}{2\sqrt{L_1 L_2}}\\
	i \frac{c|t_m|}{2\sqrt{L_1 L_2}} &  i(\Delta -G_2 \Delta x) - \kappa_2/2
	\end{pmatrix}
	\begin{pmatrix}
	\alpha_1(t) \\
	\alpha_2(t) 
	\end{pmatrix},
	\end{align}
	with $\kappa_j =  c|t_j|^2/(2L_j)$. Most importantly, we can immediately identify the off-diagonal elements as the hopping rate
	\begin{align}\label{eq:J}
	J = \frac{c|t_m|}{2\sqrt{L_1 L_2}} = \frac{c|t_m| \sqrt{-G_1 G_2}}{2\omega_0} .
	\end{align}

\end{document}